\documentclass[reprint,pre,aps]{revtex4-2}
\usepackage{graphicx}
\usepackage{amsfonts}
\usepackage{natbib}
\usepackage{xcolor}
\usepackage{amsmath}
\usepackage{enumerate}
\usepackage[colorlinks=true, linkcolor=blue, citecolor=blue, urlcolor=blue]{hyperref}
\usepackage{booktabs}
\newcommand{\be}{\begin{equation}}
\newcommand{\ee}{\end{equation}}
\newcommand{\bea}{\begin{eqnarray}}
\newcommand{\eea}{\end{eqnarray}}

\newcommand{\ep}{\varepsilon}

\newcommand{\mps}{\langle p^2 \rangle}

\begin{document}
\title{Quantum chaos and semiclassical behavior in mushroom billiards I: Spectral statistics}

\author{Matic Orel, Črt Lozej, Marko Robnik and Hua Yan}

\affiliation{CAMTP - Center for Applied Mathematics and Theoretical
  Physics, University of Maribor, Mladinska 3, SI-2000 Maribor, Slovenia,
  European Union}

\date{\today}

\begin{abstract}
  We study the aspects of quantum chaos in mushroom billiards introduced by Bunimovich. This family of billiards classically has the property of mixed phase space with precisely one entirely regular and one fully chaotic (ergodic) component, whose size depends on the width $w$ of the stem, and has two limiting geometries, namely the circle (as the integrable system) and
  stadium (as the fully chaotic system). Therefore, this one-parameter system is ideal to study the semiclassical behavior of the quantum counterpart. Here, in paper I, we study the spectral statistics as a function of the geometry defined by $w$, and as a function of the semiclassical parameter $k$, which in this case is just the wavenumber $k$. We show that at sufficiently large $k$ the level spacing distribution is excellently described
  by the Berry-Robnik (BR) distribution (without fitting). At lower $k$ the small deviations from it can be well described by the Berry-Robnik-Brody (BRB) distribution, which captures the effects of weak dynamical localization of Poincar\'e-Husimi functions.
  We also employ the analytical theory of the level spacing ratios
  distribution $P(r)$ for mixed-type systems, recently obtained by Yan (2025), which does not require a spectral unfolding procedure, and show excellent agreement with numerics
  in the semiclassical limit of large $k$. In paper II we shall analyze the eigenstates by means of Poincar\'e-Husimi functions. 
  
  \end{abstract}

\maketitle

\section{Introduction}
\label{sec1}

Quantum chaos (or more generally, wave chaos, regarding general wave systems) is a study of features of quantum systems corresponding to the structures of classical phase space of the
classical counterpart, especially in the semiclassical limit, or short wavelength regime \cite{Stoe,Haake,Rob2020,Rob2023Rev}. The structure of wave functions as well as of the corresponding Wigner and Husimi functions, defining the quantum phase space,
is of interest. In addition, the statistical properties of energy spectra and of other observables
reflect the semiclassical behavior of the systems. The classically integrable systems
exhibit Poissonian statistics \cite{BerTab1977,RobVeb}, while fully chaotic (ergodic) systems
obey the statistics of the random matrix theories (RMT) \cite{Cas1980,BGS1984},
of Gaussian Orthogonal Ensemble (GOE)
or Gaussian Unitary Ensemble (GUE), depending on
the existence or nonexistence of the antiunitary symetries in the system,
respectively \cite{RobBer1986,Rob1986}.
The latter statement is known as the Bohigas-Giannoni-Schmit conjecture \cite{BGS1984}),
which has been initiated by Casati, Guarnieri and Valz-Gris \cite{Cas1980}.
In between, in cases of mixed type classical
phase space, the Berry-Robnik picture applies \cite{BerRob1984,ProRob1999,Rob2023Rev},
based on the Principle of Uniform Semiclassical Condensation of Wigner
functions (PUSC) \cite{Rob1998}. 

Billiard systems are very interesting and attractive systems to study,
as their classical dynamics is easy to follow and the corresponding Schr\"odinger equation
(in this case the Helmholtz equation) is easier to solve than in the smooth Hamiltonian systems,
although - as we shall see - it also presents serious technical problems, which we have
succesfully overcome in this work. In particular, billiard systems exhibit all the general and
generic properties of other physical and model systems, and thus are convenient to study for the
purpose of general empirical analysis and testing the theoretical approaches and results.

In these two papers I and II we study the mushroom billiards introduced by Bunimovich \cite{Bunimovich2001},
which have been topics of several extensive studies \cite{Altmann2005,BarBet2007,bunimovich2012,Bunimovich2014} .
These billards have
sharply divided classical phase space, consisting of precisely one regular and one fully
chaotic (ergodic) component. Moreover, the invariant measure of the chaotic (and/or regular)
component can be analytically calculated. Therefore, this family of billiards is ideal to
study semiclassical theories, meaning the behavior in the semiclassical limit of sufficiently small
effective Planck constant, or equivalently of small wavelength, or sufficiently high
wavenumber $k$, in order to demonstrate the above-mentioned Berry-Robnik picture, in
the asymptotic limit $k\rightarrow \infty$, and the deviations at smaller $k$ due to the
quantum (or dynamical) localization, effects due to the bouncing ball modes, scars  and
tunneling effects.

In this paper I we concentrate on the statistical properties of the energy spectra, and confirm
that in the semiclassical limit of sufficiently large $k$ the Berry-Robnik statistics (for
our case of one regular and one chaotic region) applies, as shown for the level spacings
distribution. Moreover, we also demonstrate that the very recently developed theory
by Yan (2025, \cite{Yan2025}) of the level spacings ratio distribution applies in the
asymptotic limit, for mixed-type systems. Furthermore, we notice at smaller $k$ small deviations
from the behavior in this deep semiclassical limit, which are attributed mainly to the weak dynamical, or quantum, localization of the Husimi functions (in this case Poincar\'e-Husimi functions).
These effects are well captured by the Berry-Robnik-Brody distribution. The structure of the eigenstates and the corresponding Poincar\'e-Husimi functions will be the topic of paper II. 

  This paper is organized as follows: In Sec. \ref{sec2} we define the geometry of the mushroom billiards, describe the classical dynamics, present the relative Liouville measure of the chaotic component and discuss the classical transport time. In Sec. \ref{sec3} we define the Schr\"odinger equation (which is the Helmholtz equation) and present an extensive analysis of various numerical techniques, their advantages and shortcomings (detailed
  in Appendix \ref{appA}). In Sec. \ref{sec4} we present the statistical analysis of the level spacings distribution
  for the entire spectrum, while in Sec. \ref{sec5} we analyze the level spacings ratio distribution, employing the recent theory from Ref. \cite{Yan2025}. In Sec. \ref{sec6} we discuss the effects of localization as reflected in the statistics of the separated chaotic part of the spectrum. In Sec. \ref{sec7} we discuss the results and conclude.

\section{The mushroom billiards}
\label{sec2}

\subsection{Geometry and classical dynamics}

\begin{figure}
 \begin{centering}
   \includegraphics[width=9cm]{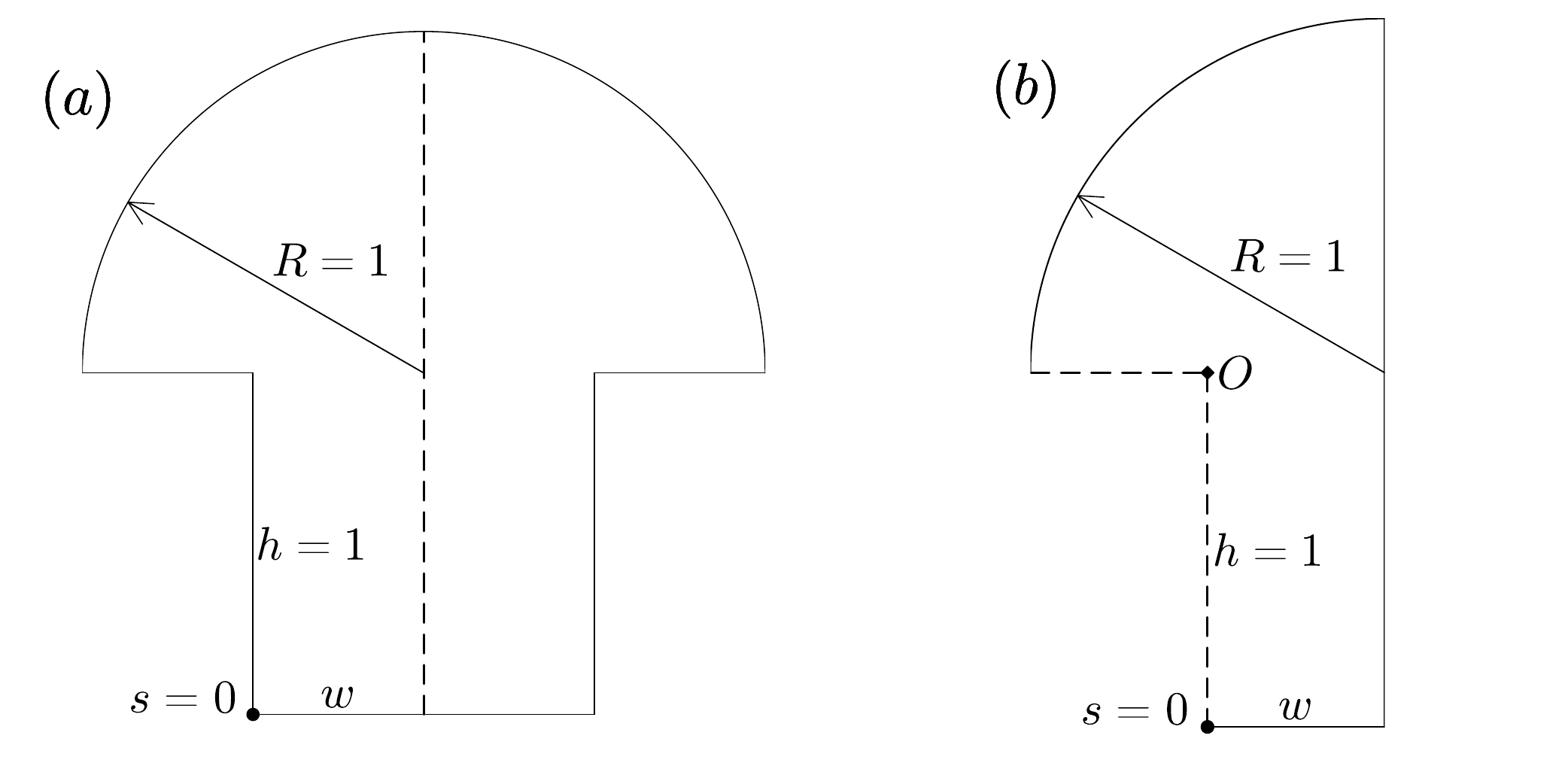}  
   \par\end{centering}
 \caption{The geometry of the mushroom billiard. For the classical
   dynamics we use the full billiard (a), while for the quantum calculations
   we use the desymmetrized billiard (b) to obtain the odd parity solutions of (a)
   with respect to the
   symmetry line (dashed). The Poincar\'e-Birkhoff coordinates are $(s,p)$,
   where $s$ is counted anti-clockwise from the origin $s=0$, while $p$ is the
   sine of the reflection angle. The origin $O$ of
   the Fourier-Bessel basis for the Vergini-Saraceno method is placed 
   at the corner in such a way that
 the boundary condition on the dashed segment of (b) is satisfied by the construction.}
\label{geom}
\end{figure}

In Fig. \ref{geom} we show the geometry of the mushroom billiard as we use it in this paper.
The boundary consists of a unit semicircle   $R=1$,
connected to the stem of unit length $h=1$, and width $2w$. For the
classical dynamics we study the full billiard (a),
while for the quantum mechanics we use half billiard (b),
thereby restricting the solutions to the odd symmetry class with respect to the reflection symmetry.
Thus, the width of the stem of the half billiard is $w$.
We use the Poincar\'e-Birkhoff coordinates in the phase space (of the bounce map), namely
$s$, the arclength around the billiard boundary starting at the point $s=0$, and counted
in the anti-clockwise direction. The conjugate momentum is $p=\sin \alpha$, where
$\alpha$ is the reflection angle  \cite{Berry1981}. As explained in the next section \ref{sec3}, we mainly use the Vergini-Saraceno method to solve the Schr\"odinger equation (Helmholtz equation), specifically the Fourier-Bessel basis (\cite{VerSar1995,BarBet2007}, designed in such a way that
the Dirichlet boundary condition is satisfied, by construction, on the dashed segment of the boundary in Fig. \ref{geom}(b).

As it is well known, the mushroom billiard has exactly one regular and exactly one chaotic
component in its phase space $(s,p)$, with a smooth boundary between them, in contradistinction
to the KAM systems,
which possess infinite hiererchical regular and chaotic structures. This property is
the key aspect convenient to study quantum chaos in mixed-type systems \cite{BerRob1984,Rob1998,Rob2023Rev},
which is the main objective of this paper. All orbits that live only in the cap (bounded by
the semicircle) are
regular, while those orbits that enter the stem are chaotic---except for a set of marginally unstable periodic orbits (MUPOs), which form the boundary between the regular and chaotic regions. These orbits belong to the chaotic region but, intriguingly, never enter the stem. They are tangent to the semi-circle connecting the two corners, congruent with the unit semi-circle, having the radius equal to $w$ and the trace of the monodromy matrix has absolute value 2. They give rise to the stickiness effects, and play a role in the structures of quantum
eigenstates. The family parameter $w$ takes us from the entirely integrable case (semi-circle, $w=0$) to the entirely chaotic and ergodic case (half-stadium, $w=1$), leaving the mixed-type regime in between $0<w<1$. It is important to note
that in the stem we also have a continuum of MUPOs, the so-called
bouncing ball modes, exactly as in the stadium, which  give rise to stickiness in the chaotic region. The plentiful unstable periodic orbits embedded in the ergodic chaotic region would give rise to the quantum scars in the sense of Heller \cite{Heller1984}, as observed in the Poincar\'e-Husimi functions of the chaotic eigenstates, to be discussed in paper II.

\begin{figure}
 \begin{centering}
   \includegraphics[width=0.95\linewidth]{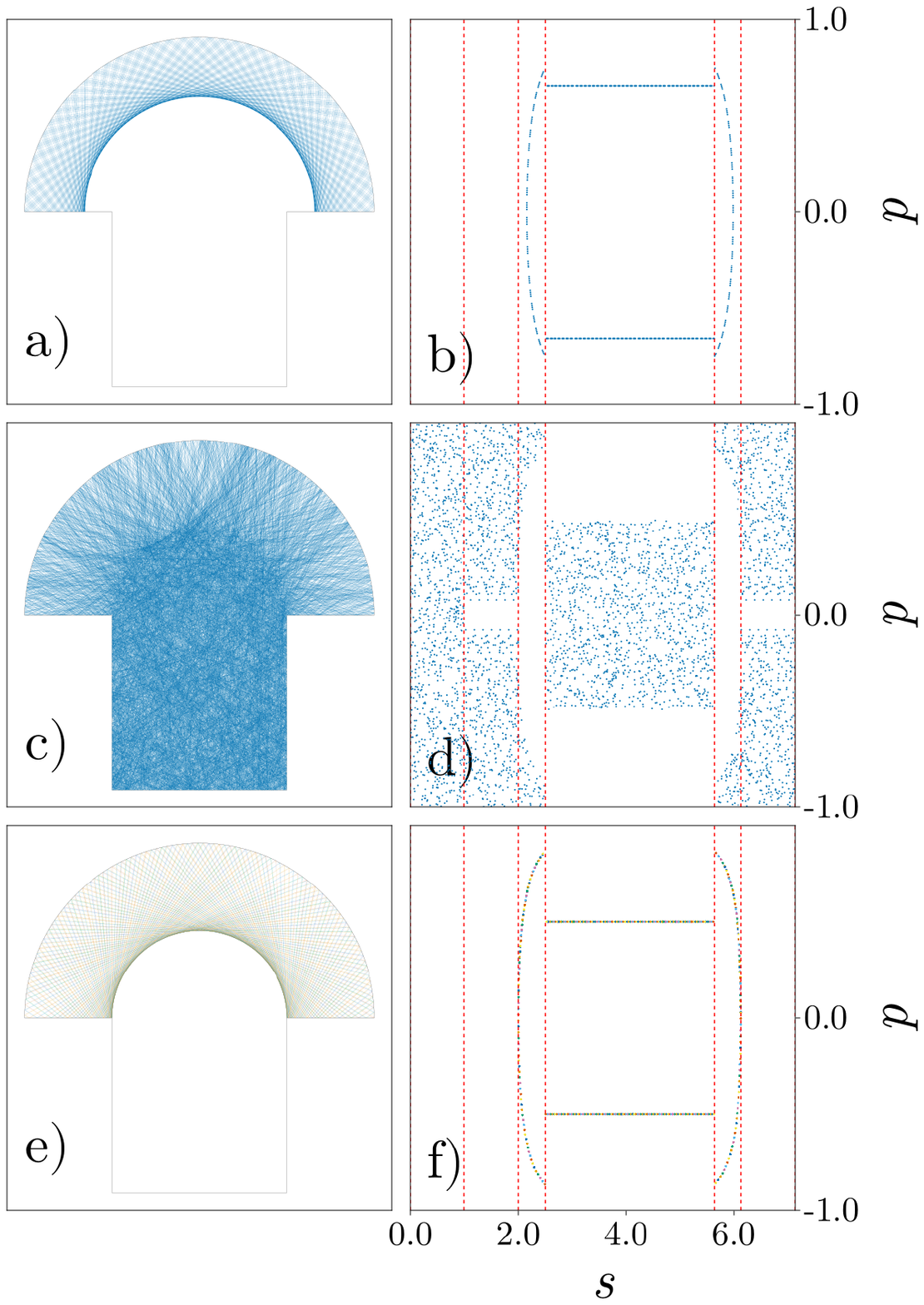}  
   \par\end{centering}
 \caption{A typical regular orbit in configuration space (a) and in the
   phase space (b), and a typical chaotic orbit correspondingly in (c,d).
 We also show examples of MUPO in (e,f). The vertical dashed lines signify the endpoints of individual segments forming the full boundary with arclength s following Fig. \ref{geom}.}
\label{orbits}
\end{figure}

In Fig. \ref{orbits} we show examples of typical regular and typical chaotic orbit, both
in the configuration space
and in the phase space, after 4000 collisions. We also show members of the family of
the MUPOs, plotted for 20 initial conditions, where we clearly see that they are all tangent
to the semicircle of radius $w$.

\label{sec2.1}

\subsection{The relative Liouville volume measure of the chaotic part of the phase-space}
\label{sec2.2}
\begin{figure}
 \begin{centering}
   \includegraphics[width=0.95\linewidth]{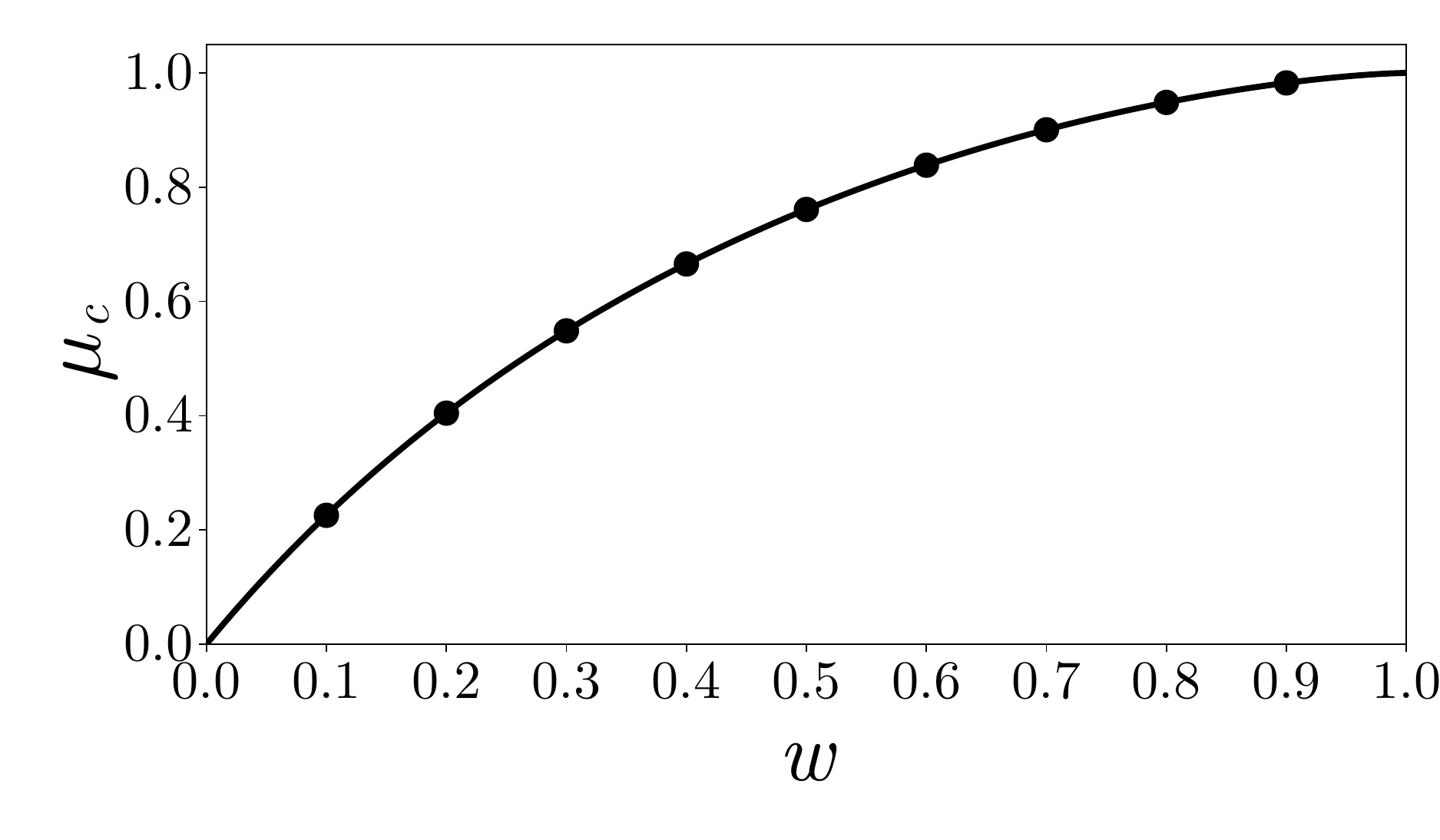}  
   \par\end{centering}
 \caption{The relative fraction of the Liouville volume of the chaotic component
   (Berry-Robnik parameter) $\mu_c(w)$ according to the general
 equation (\ref{mucl}), with $R=h=1$.}
\label{figmuc}
\end{figure}

A mixed-type system, ranging from sharply divided to hierarchical phase space, exhibits both regular and chaotic orbits depending on the initial conditions. An initial condition in the phase space belongs to the chaotic sea if it generates a
chaotic orbit whose largest Lyapunov exponent $\lambda$ is positive \cite{BouSko2012},
or the smaller alignment index (SALI) exhibits exponential decay over time. In billiard systems, SALI typically decreases exponentially with the number of collisions $n$, as $\propto \exp (-\lambda n)$. One of the key parameters characterizing a mixed-type Hamiltonian system is the relative invariant measure $\mu_c$  of the chaotic part of
the phase space. This is simply the relative Liouville volume of the
chaotic sea (more precisely, of the dominant chaotic region)
on the energy hypersurface. (It should be emphasized that $\mu_c$
is different from the relative area of
the chaotic region in the surface of section. See e.g. Refs. \cite{meyer1986,BatRob2010}.)
In the general mixed-type systems having a KAM hierarchical structure of
infinitely many regular islands surrounded by a chaotic sea, an analytic calculation
of $\mu_c$ is usually impossible.  In the mushroom billiards this difficulty
is removed, and $\mu_c$ can be calculated exactly analytically, resulting in the
closed form (see e.g. Ref. \cite{BarBet2007}):

\be \label{mucl}
\mu_{c}= 1 - \frac{\frac{R^2}{2}
  (\arccos(\frac{w}{R})-\frac{w}{R}) \sqrt{1 - (\frac{w}{R})^2})}{wh + \frac{\pi R^2}{4}}.
\ee
In our case $R=h=1$. In Fig. \ref{figmuc} we plot the graph $\mu_c(w)$.

In the analysis of the energy spectra of mixed-type systems $\mu_c$ is the crucial parameter,
also called Berry-Robnik parameter introduced in Ref. \cite{BerRob1984},
which shall be used in Sec. \ref{sec4}.

\begin{figure*}
 \begin{centering}
   \includegraphics[angle=0,width=16cm]{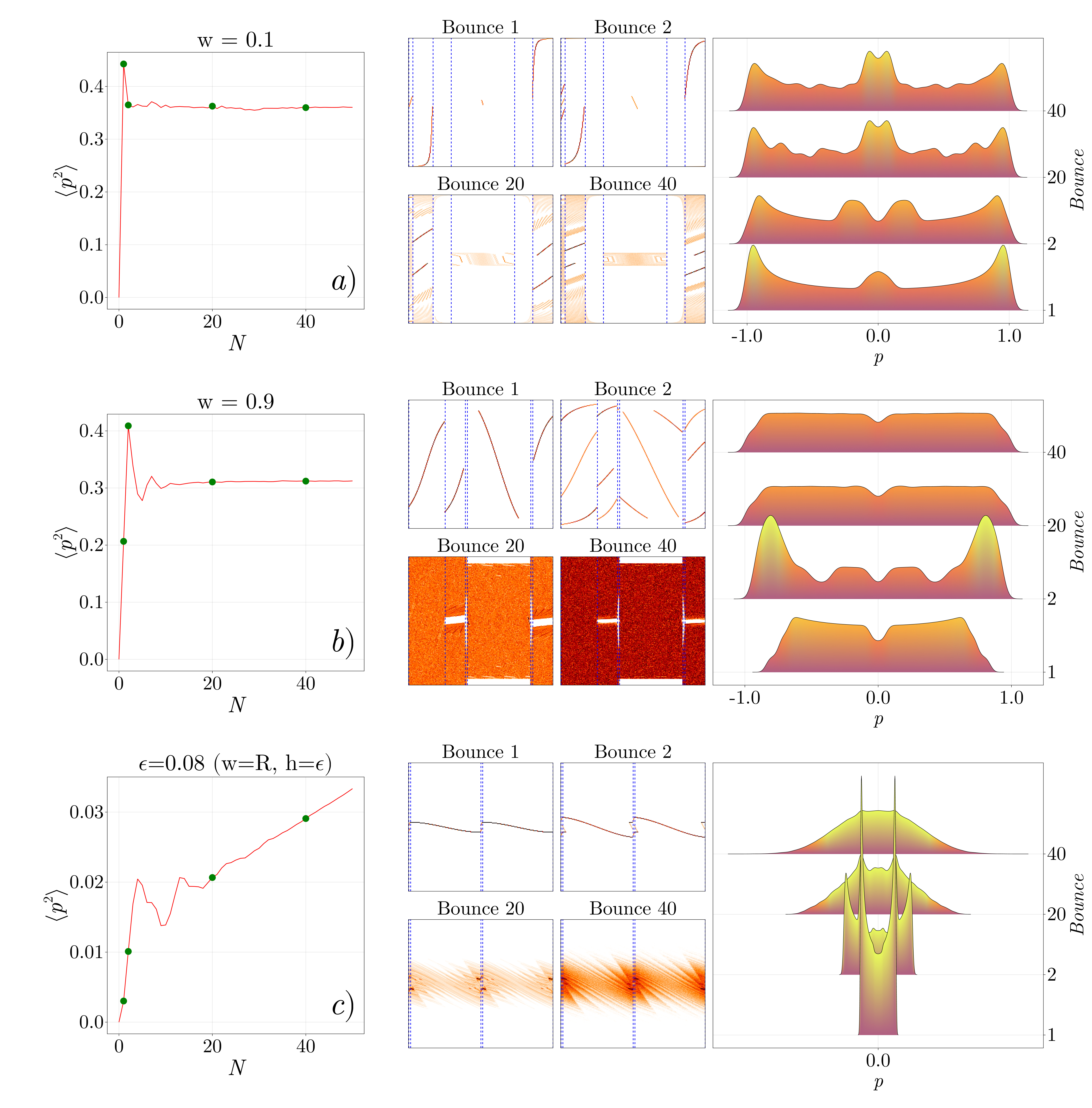}  
   \par\end{centering}
 \caption{In (a) and (b) we show $\mps$ as a function of $N$ for $w=0.1$ and $0.9$, respectively.
   In (c) we show the result for the stadium billiard with radius $1$ and
   $\ep=0.08$ (the length of the straight line connecting the two semicircles).
   The difference in behavior is evident. The accompanying plots
   show the structure of the phase space of the ensemble $p=0$ after 1, 2,20, and 40 bounces,
   together with the belonging distribution of $p$. Thus, a different measure
 of the ensemble spreading must be used.}
\label{figNTp2}
\end{figure*}

\subsection{The classical transport time}
\label{sec2.3}

In every quantum Hamiltonian system with a purely discrete energy spectrum, the Heisenberg time,
defined by $t_H = 2\pi\hbar/\langle \Delta E \rangle$, with $\langle \Delta E \rangle$
being the mean energy level spacing, is an important time scale.
As has been explained in Ref. \cite{BatRob2013A} the 
time evolution of quantum states, starting e.g. with a wave packet (coherent state),
in the quantum phase space (Wigner function or Husimi function)
follows the classical diffusion up to the Heisenberg time $t_H$ (also called break
time by Boris Chirikov and Giulio Casati), because 
the discreteness of the Hamilton operator is resolved only at times longer than Heisenberg time.
Therefore, the quantum diffusion stops when reaching (approximately) the Heisenberg time, and
if the classical diffusion is slow, such that the classical transport or diffusion time $t_T$ is
longer than the Heisenberg time, we observe localization of the asymptotic final
state, just due to the destructive interference effects. In addition to this time dependent picture,
we typically observe localized stationary eigenstates, localized in the quantum phase space,
represented by the Wigner or Husimi functions, which do not occupy the entire
available classical chaotic region. The localization criterion is expressed in
terms of the parameter $\alpha=t_H/t_T$: If $\alpha <1$ we find localization, while
if $\alpha >1$ the extendedness of the eigenstates in the phase space appears. However, as demonstrated in previous works \cite{BatRob2013A,BatRob2013B}, this transition is not
an abrupt (discontinuous) one, but rather smooth, taking place generally over an interval $\alpha$  of one decadic
order of magnitude around $\alpha=1$. Moreover, in the strict semiclassical limit
$\alpha \rightarrow \infty$, because $t_H \propto \hbar^{(1-f)} \rightarrow  \infty$,
as $\hbar \rightarrow 0$, where $f$ is the number of degrees of freedom. In 2-dimensional
billiards $f=2$.

In billiards \cite{BatRob2013A}, the parameter $\alpha$
is estimated as $\alpha = \frac{{\cal L} k}{\pi N_T}$, where $k$ is
the wave number (using the units defined in Sec. \ref{sec3}), ${\cal L}$ is the
length of the billiard boundary, in our case  ${\cal L} = 4+\pi/2\approx 5.57$, while $N_T$ is the
classical transport time in terms of the number of collisions.
Thus, $\alpha \approx  1.77\; k/ N_T$. To calculate $N_T$, usually,
one defines an ensemble of initial conditions, e.g. taking a uniform distribution on the
line $p=0$. Then the ensemble evolves, the mean value $\langle p^2\rangle$ is calculated at each collision $N$, following a diffusion law $\mps \propto N$, before
the boundary effects become apparent and important.
$N_T$ is then a typical time to reach its saturation value.
For details, see Refs. \cite{BatRob2013A,LozRob2018}.If the dynamical billiard is slowly ergodic (like for example the small stadia) the diffusive process in momentum space may be described by an inhomogeneous normal diffusion model \cite{LozRob2018}. The saturation is exponential, and thus a natural timescale may be defined. the diffusive regime is only established after the initial scrambling time, where the correlations between the initial conditions are lost. The scrambling time may indeed be longer than $N_T$ and consequently the distribution is spread over the available phase space before the diffusive process may start in earnest. This is very common in strongly chaotic and mixed-type billiards, especially when the boundary contains some discontinuities. This can clearly be seen in Fig. \ref{figNTp2} where in cases a) and b) two different mushroom geometries produce a non-diffusive momentum spreading while in the stadium with small $\epsilon$ c) its spreading follows normal diffusion after the initial scrambling of the initial conditions (the distribution assumes a Gaussian shape and continues to spread). Therefore, Eq. 19 as derived in \cite{LozRob2018} does not hold in our case, and we require a different criterion for the determination of $N_T$.

\begin{figure}
 \begin{centering}
   \includegraphics[width=1.0\linewidth]{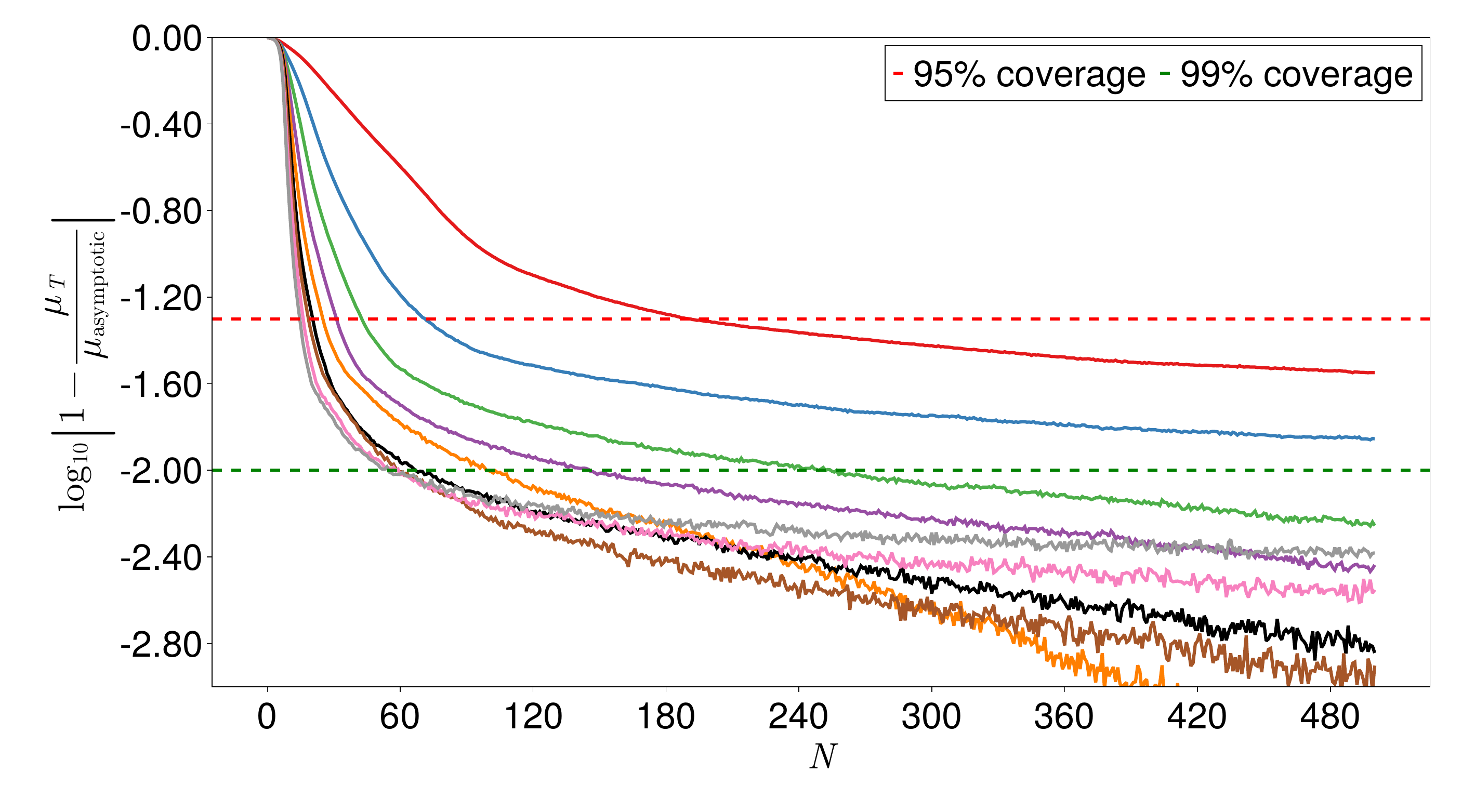}  
   \par\end{centering}
 \caption{The relative area of the occupied chaotic component, with
   respect to the asymptotic value, $\mu_T(N)/\mu(\infty)$, starting
   with an uniform ensemble of $10^6$ initial conditions located at
   $p=0$, with $R=h=1$ and 9 different  $w$ (top). For the calculation of
   the relative chaotic area $\mu_T$ a grid of $450 \times 450$ cells is used.
   The classical transport time
   $N_T$ is estimated as time at which this quantity reaches 95\% of
   its asymptotic value. Thus, $N_T$ is in the range between 10 and 180 (bottom),
   for $w=0.9$ to $w=0.1$ in equal steps of $0.1$.}
\label{figmucN}
\end{figure}

We chose the filling ratio of the occupied chaotic region $\mu_T$ relative
to its asymptotic value as the criterion for determining $N_T$. This is shown in Fig. \ref{figmucN}, where 95\% and 99\% are used as the threshold. We see that 99\% is hardly reached
for  $w=0.1$ and $w=0.2$, 
while 95\% gives a reasonable estimate, showing that $N_T$ is in the range
$10$ and $180$, when $w$ goes from $0.9$ to $0.1$ in equal steps $0.1$.
Further details regarding the aspects of the transport time and its
relevance for the localization of eigenstates in the phase space will be discussed
in Sec. \ref{sec6}.

\section{The Helmholtz equation and the numerical techniques}
\label{sec3}

The quantum billiard ${\cal B}$, in our case the half-mushroom billiard,
depicted in Fig. \ref{geom}(b), is described
by the stationary Schr\"odinger equation, in the chosen units
($\hbar^2/2m=1$) it is the simple Helmholtz equation

\be \label{Helmholtz}
\Delta \psi + k^2 \psi =0
\ee
where the Dirichlet boundary conditions  $\psi|_{\partial {\cal B}}=0$
must be satisfied. The energy is $E=k^2$. 

The {\em mean number of energy levels}
${\cal N} (E)$ up to $E=k^2$ is determined asymptotically exactly
by the celebrated Weyl formula (with perimeter corrections),
taking into account the Dirichlet boundary conditions, namely

\be \label{WeylN}
{\cal N} (E) = \frac{{A}\;E}{4\pi} - \frac{{L}\;\sqrt{E}}{4 \pi} + c,
\ee
where $c$ are small constants due to the corners and the
curvature of the billiard boundary. For the density of levels
$d (E) = d{\cal N}/dE$ they are irrelevant,

\be \label{Weylrho}
d(E) = \frac{{A}}{4\pi} - \frac{{L}}{8 \pi \sqrt{E}}.
\ee
In a previous paper \cite{LLR2021B} we have demonstrated, in agreement with the
theoretical predictions by Steiner \cite{Steiner1994,SteinerPRL1994}, that
the fluctuations of the number of energy levels (mode-fluctuations)
around the mean value of Eq. (\ref{WeylN}) grow with $k$. Their variance
increases linearly with $k$ in integrable systems, while it grows as $\log k$
in ergodic chaotic systems. For the systems with divided phase space, it is something in between, namely the variance is the sum of the variances of the regular and of the chaotic part, provided that
they are statistically independent of each other. 
In the semiclassical regime this is a valid assumption in the spirit
of the Berry-Robnik picture \cite{BerRob1984}. 
Therefore, the fluctuations at large $k$ in all cases
can be very large, as the standard deviation even diverges as
$k\rightarrow \infty$. In ergodic chaotic systems
the distribution of fluctuations is Gaussian
\cite{Steiner1994,SteinerPRL1994}. On the other hand, in Ref. \cite{LLR2021B}
we have shown that it is nearly Gaussian,
even in the integrable, as well as in mixed-type systems.

\begin{table*}[ht]
\centering
\begin{tabular}{@{}l p{5.5cm} p{5.5cm}@{}}
\toprule
\textbf{Method} & \textbf{Pros} & \textbf{Cons} \\
\midrule
Boundary Integral (Sweep), in Sec. \ref{nm1}
  & Robust and can target individual symmetries 
  & One eigenvalue per SVD (singular value decomposition); slow for full spectrum \\
Expanded BIM (Accelerated), in Sec. \ref{nm2}
  & Multiple eigenvalues per solve via GEP (generalized eigenvalue problem), can target individual symmetries 
  & Non-Hermitian GEP; poor conditioning at high $k$ \\
Decomposition (Sweep), in Sec. \ref{nm3}
  & Enforces boundary vanishing directly in GEP by construction; avoids area integrals 
  & One root per GEP  solve; requires two boundary matrices to be constructed \\
Particular Solutions (Sweep), in Sec. \ref{nm4}
  & Replaces area integrals with interior sampling 
  & Needs many points; sampling error possible especially with localized states\\ 
Vergini–Saraceno (Accelerated), in Sec. \ref{nm5}
  & Multiple eigenvalues per diagonalization, offering high efficiency and stability 
  & Breaks down quickly for non-convex domains without tailored basis \\
\bottomrule
\end{tabular}
\caption{Comparison of sweep-based and accelerated methods for eigenvalue computation. Sweep based methods require checking every trial eigenvalue if it is correct while accelerated methods give many true eigenvalues in a small interval.}
\label{tab:simple_methods}
\end{table*}

Our numerical method to compute the eigenfunctions is based on the Vergini-Saraceno scaling method
\cite{VerSar1995,LozejThesis}, explained in detail in Sec. \ref{nm5} and  available as part of a numerical package \cite{QuantBill}. This method finds all pairs of eigenvalue and base expansion coefficients $(k,\mathbf{x})$ in a small interval $k \in [k_{0}-\delta k, k_{0} + \delta k]$. Since the error of the method has an upper bound determined by the interval width $\delta k$, which scales as $|k-k_{0}|^3$, it's width must be chosen as small as possible without sacrificing redundant computational effort. Our choice was $\delta k = 0.01$, together with the number of collocation points on the boundary scaling as $N_{boundary} = {7 k \mathcal{L} }/{2 \pi}$ and the number of basis functions scaling as $N_{basis} = {4 k \mathcal{L} }/{2 \pi}$. Calculation of the semiclassical basis size using the turning point rule \cite{Barnett2000}

\begin{align}
    J_{\frac{2i}{3}} (kr) \sim 0 \quad \text{for} \quad kr<\frac{2i}{3},
\end{align}
and using the fact that the largest distance inside the billiard $R=\max_{(x,y)\in\partial\Omega}\sqrt{x^2 + y^2} = \sqrt{1+w^2}$ gives the semiclassical basis size $N_{sc} = 3k\sqrt{1+w^2}/2$ showing that for all widths $w$ our choice for basis size is more than adequate. We computed up to  $1.3\times 10^5$ eigenstates
in the PH representation for each billiard with given $w$.
However as noted above the precision of the computed energy levels decreases with the
distance from the center of the energy interval. Thus, even after careful
comparison of the levels in overlapping energy intervals, errors in the
accumulation of levels still occur, and some levels are lost.
This can be detected by the drift of the locally averaged value of numerical
${\cal N} (E)$ as compared with the Weyl ${\cal N} (E)$ of equation (\ref{WeylN}). The number
of missing levels was never larger than a few per $10000$ levels and due
to the overall large number of eigenstates this should have very little
effect on the statistical results. The latter aspect has been verified by
artificial defects of spectra, by simply deleting/removing exactly $0.1\%$ randomly
chosen levels, and making sure that the spectral statistics does not change
by doing this.

We must mention that our numerical computations have been based
on several competing methods, whose results and their accuracy
as well as their computational efficiency (algorithmic efforts) have been
thoroughly cross-checked. Apart from the Vergini-Saraceno method using the corner-adapted 
Fourier-Bessel basis as the ultimately primary (superior) method,
we have used the following alternative methods summarized in Table \ref{tab:simple_methods}: elementary boundary integral method
\cite{BW1984}, expanded boundary integral method \cite{VebProRob2007},
decomposition method \cite{Barnett2000}, particular solutions method \cite{BarBet2007},
conformal mapping diagonalization technique \cite{Rob1984} and the finite element
method.  We describe the details of the methods in Appendix \ref{appA}.
After many careful checks it turned out that for our specific case of
the mushroom billiard, the corner-adapted Fourier-Bessel version of the Vergini-Saraceno
method is superior.

\section{The statistical analysis of the energy spectra: Level spacings distribution}
\label{sec4}

The main purpose of this paper, studying the energy spectra of the mushroom billiards
in the semiclassical limit,
is threefold: (i) To demonstrate the applicability of the Berry-Robnik level spacings
distribution in mixed-type system \cite{BerRob1984}, (ii) to demonstrate the applicability of the
recent theory by Yan \cite{Yan2025} of the level spacing ratio distribution in
the mixed-type case, and (iii) to demonstrate the effects of weak phase space
localization in the Poincar\'e-Husimi functions, at lower energies, below the
strict semiclassical limit, due to the effects of finite $N_T$, characterized by
$\alpha \leq 1$.

\begin{figure*}
 \begin{centering}
   \includegraphics[width=0.9\linewidth]{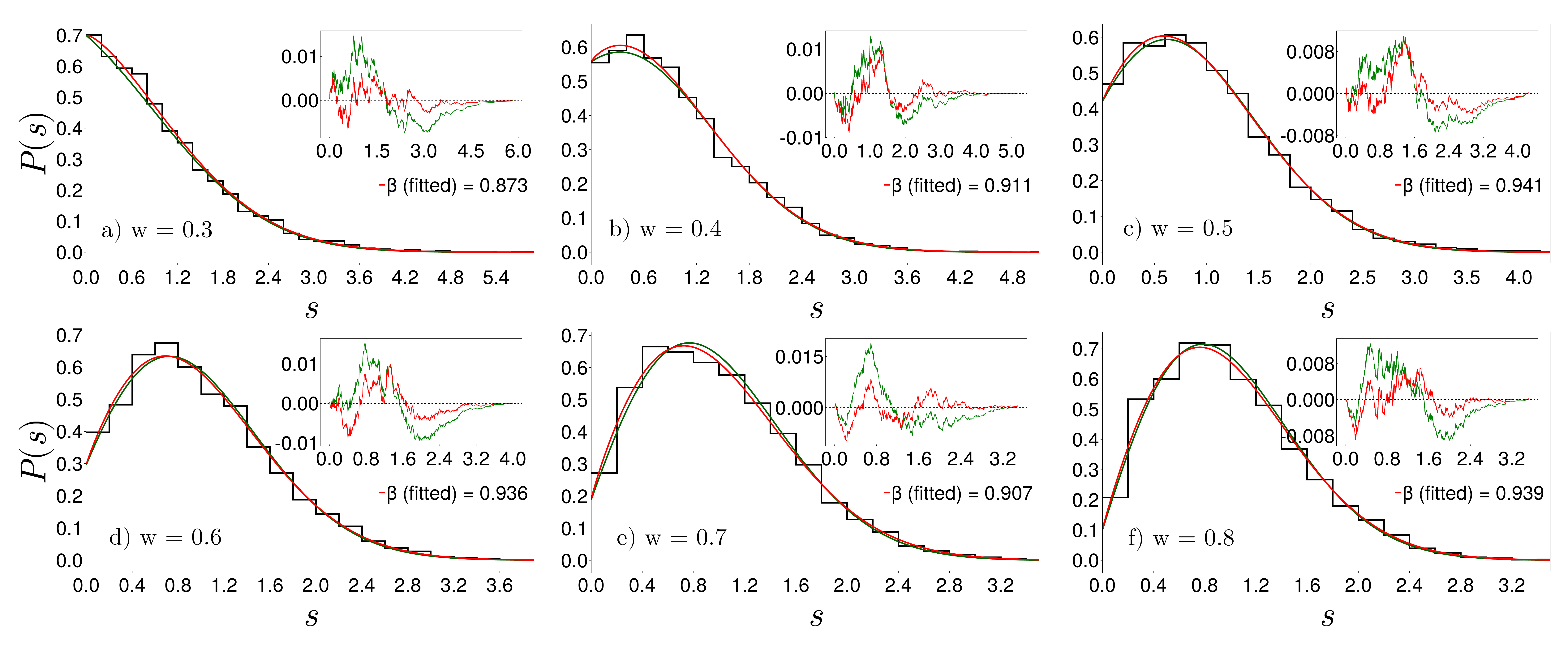}  
   \par\end{centering}
 \caption{Numerical level spacing distribution of the spectrum stretch $k\in [970,1000]$
 compared with the BR theory \cite{BerRob1984} Eq. (\ref{BRPS})
   using the exact analytical
   value of $\mu_c$, $\rho (reg) = 1 - \mu_c$, for 6 geometries $w=0.3-0.8$ in equal steps of $0.1$,
   from (a) to (f). The green solid line represents the BR distribution and the red solid one the BRB distribution. We also show best fitting BRB distribution with the analytical $\mu_c$ and best fitting
   value of $\beta$. In the inset we plot the differences between the numerical CDF and BRB CDF (red) and between the numerical CDF and BR (green). It is clearly seen that
   theoretical BR and best fitting BRB distributions are very close to each other, as the localization effects
 are minor.}
\label{figPSallhigh}
\end{figure*}

First we show in Fig. \ref{figPSallhigh} the level spacings distribution for the spectrum
within the range $k\in \lbrack 970, 1000 \rbrack$, obtained after the unfolding
using the Weyl formula - equations (\ref{WeylN},\ref{Weylrho}).
The level spacings distribution $P(s)$ is generally equal to the second derivative
of the gap probability $E(s)$ of having no level on an interval of length $s$, 
$P(s) = d^2E(s)/ds^2$. In the absence of correlations between the regular
and chaotic levels of relative fraction
$\mu_r$ and $\mu_c=1-\mu_r$, correspondingly, the gap probability $E(s)$ 
is obviously just the product
of the corresponding gap probabilities $E_r(s)$ and $E_c(s)$ with the corresponding
weights in their arguments, so that

\be \label{gapS}
E(s) = E_r(\mu_r s) E_c(\mu_c s).
\ee
For a regular spectrum the level statistics is Poissonian, $E_P(s)=\exp (-s)$,
$P(s)= \exp (-s)$,  while
for the fully chaotic case the GOE of random matrix theory applies, well approximated
by the Wigner distribution, if there is no phase space localization,

\be \label{PSWigner}
P_W (s) = \frac{\pi s}{2} \exp \left( - \frac{\pi s^2}{4}\right),
\ee
so that
\be \label{ESWigner}
E_W (s) =  1 - \rm{erf} \left( \frac{\sqrt{\pi} s}{2}\right) = \rm{erfc} \left( \frac{\sqrt{\pi} s}{2}\right),
\ee
where  $\rm{erf}(x) = \frac{2}{\sqrt{\pi}}\int_0^x e^{-u^2}\, du$ is the error integral and $\rm{erfc}(x)$ its complement,
i.e. $\rm{erfc}(x) = 1 - $$\rm{erf}(x)$.
Consequently, in absence of phase space localization of the chaotic eigenstates, we immediately derive the
Berry-Robnik (BR) level spacing distribution  \cite{BerRob1984}, 

\begin{figure*}
 \begin{centering}
   \includegraphics[width=0.9\linewidth]{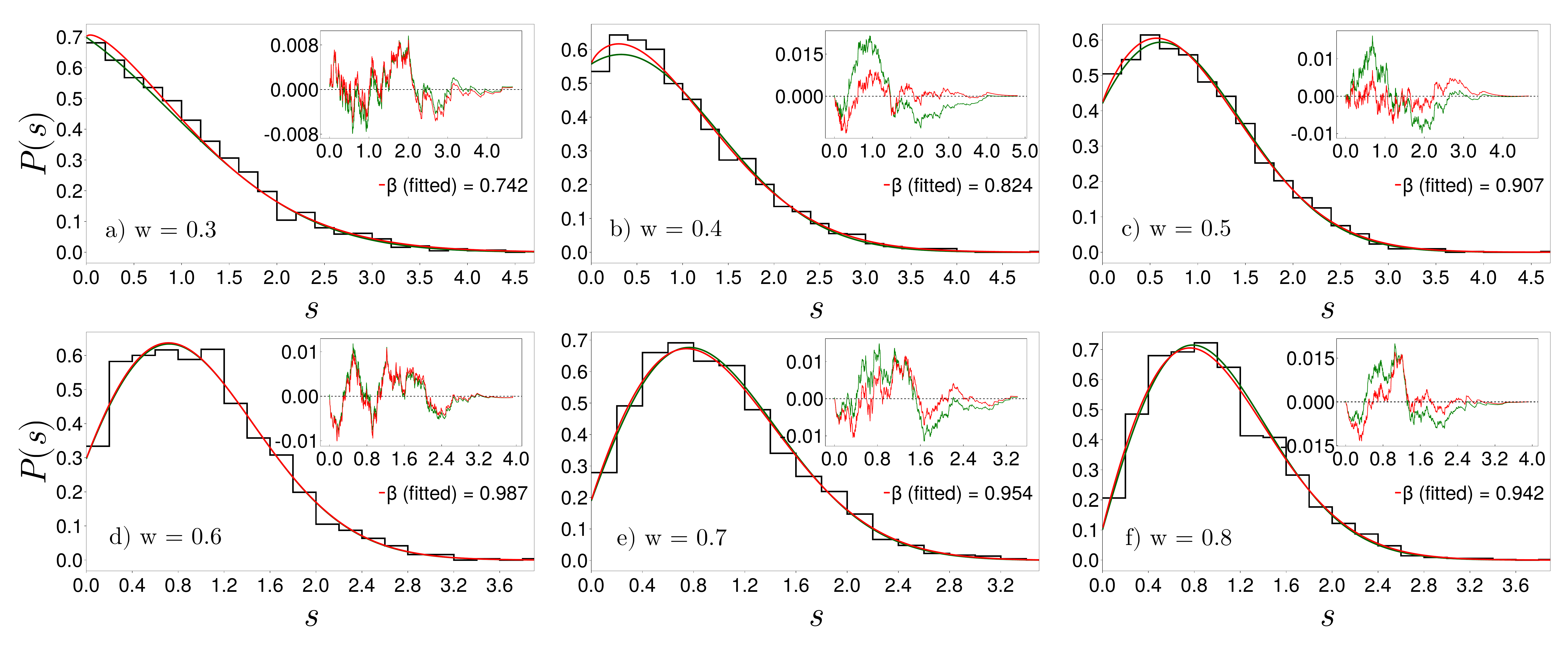}  
   \par\end{centering}
 \caption{The same as in Fig. \ref{figPSallhigh}, but for $k\in \lbrack 5, 150\rbrack$,
   showing that even at low energies $k^2$ the effects of localization are minor,
   as BR and BRB curves almost overlap. }
\label{figPSalllow}
\end{figure*}

\bea \label{BRPS}
P_{BR}(s) & = & e^{-\mu_r  s}   e^{-\frac{\pi \mu_c^2s^2}{4}}  \left(2\mu_r\mu_c \nonumber
    + \frac{\pi\mu_c^2s}{2}\right) \\
      & + & \mu_r^2 \; e^{-\mu_r  s} {\rm erfc}\left(\frac{\sqrt{\pi}\mu_c s}{2}\right).
\eea

A generalization of the above formula is necessary for low-lying eigenstates where some
localization phenomena occur, due to various mechanisms,
such as dynamical/quantum localization, scars, bouncing ball modes,
tunneling effects, and stickiness effects. It has been widely demonstrated  that
the level spacing distribution of quantum localized fully chaotic eigenstates
\cite{BatRob2010,BatRob2013A,BatRob2013B,BatManRob2013,ManRob2013,LLR2021,LLR2021A,LLR2022}
is well captured
by the Brody distribution \cite{Bro1973,Bro1981},
\bea \label{PSB}
P_B(s) &=& C_1 s^{\beta} \exp (-C_2s^{\beta+1} ), \nonumber\\
C_1 &=& (\beta+1)C_2,
\;\;\; C_2 = \left( \Gamma \left( \frac{\beta+2}{\beta +1} \right)\right)^{\beta+1}. 
\eea
from which the corresponding Brody gap probability follows
\be \label{EB}
E_B(s) = \frac{1}{\Gamma \left(\frac{1}{\beta+1}\right)}
Q \left(\frac{1}{\beta+1}, \left( \Gamma \left( \frac{\beta+2}{\beta +1} \right) s\right)^{\beta+1} \right),
\ee
where $Q(a,x)$ is the incomplete Gamma function:
\be  \label{incomGamma}
Q(a,x)  = \int_x^{\infty} t^{a-1}e^{-t}\,dt.
\ee
Upon replacing $E_W(s)$ by the Brody gap probability $E_B(s)$ in the formula (\ref{gapS}), we
obtain the Berry-Robnik-Brody (BRB) distribution, which now captures both effects, the divided
phase space through $\mu_c$ and the localization of the chaotic part through $\beta$.

As is well known, the limiting cases are $\beta=0$ (Poisson distribution) and $\beta=1$ (Wigner distribution).
It is our aim to show the applicability of the BR distribution (\ref{BRPS}) in the deep semiclassical limit.
This is demonstrated for high-lying states in Fig. \ref{figPSallhigh}. We compare the numerical data for
the states in the interval $970\le k\le 1000$ for 6 geometries $w=0.3-0.8$ in equal steps $0.1$. The data histogram
is compared to the theoretical BR distribution using the exact analytical value for the $\mu_c$ parameter
given in equation (\ref{mucl}), meaning that so far there is no fitting of data, but comparison of
the numerics with the BR theory.
We see statistically significant perfect agreement. However, we also show the best fitting BRB
distribution, by determining $\beta$ by fitting while keeping $\mu_c$ exact. It is evident that the two
distributions, BR and BRB are very close to each other, clearly demonstrating that the effects of
quantum localization are very weak, which is in line with the
large value of the parameter $\alpha=1.77\;k/N_T$, as $k\approx 1000$ while $N_T$
is in the range between 10 and 40,
for $w=0.8$ to $w=0.3$, in equal steps of $0.1$ (see Fig. \ref{figmucN}). 

This picture will be further corroborated below by the analysis of
spectra in different perspectives. In order to demonstrate that the localization effects are minor also at low-lying states, we show the
similar plot in Fig. \ref{figPSalllow}. It must be emphasized that we have deliberately excluded the cases of $w=0.1$ and $w=0.2$, as in these two geometries the classical dynamics is more complex, exhibiting several  time scales related to the
bouncing ball modes (in the stem) and stickiness around the marginally unstable orbits (MUPO).
Therefore, the analysis of the classical dynamics and quantum eigenstates of these two cases
will be presented in a
separate paper.


\section{The statistical analysis of the energy spectra: Level spacing ratio distribution}
\label{sec5}

The level spacing ratio distribution $P(r)$ has been introduced by Oganesyan and Huse \cite{OgaHus2007}.
It has the advantage, when compared to the level spacing distribution, that it does not require a
spectral unfolding procedure, thereby being more robust. The level spacing ratio is defined
as

\be \label{rk}
r_n = \min \left( \delta_n, {1}/{\delta_n} \right),
\ee
where $\delta_n = d_{n+1}/d_n$ with $d_n= E_{n+1}-E_n$ being the spacing between two nearest eigenenergies.
The spacing ratios distributions for the Wigner distribution are derived from the joint distribution of two consecutive spacings \cite{AtaBogGirRou2013,AtaBogGirVivViv2013}, 
derived by using $3\times 3$ Gaussian random matrices and employing the Wigner surmise,
\begin{align}
  \label{eq:joint-chaotic}
  p(s,t)= A(s+t)st \ e^{-B(s^2+st+t^2)},
\end{align}
where $A$ and $B$ are normalization constant. A further integration gives
\begin{align}
\label{PrPoissonWigner}
  p(r)&=\int_0^\infty ds dt \ p(s,t)\delta (r-\min(
    \frac{s}{t},\frac{t}{s}))\nonumber\\
  &=2\int_0^\infty  p(s,rs)sds = \frac{27}{4}\frac{r+r^2}{(1+r+r^2)^{5/2}}.
\end{align}
 It is an excellent approximation to the exact GOE result obtained for infinitely dimensional
real symmetric Gaussian random matrices (GOE), as tested numerically \cite{AtaBogGirRou2013,AtaBogGirVivViv2013}. 
For integrable systems, given the Poisson process of energy levels, the joint distribution of consecutive spacings and spacing ratios distribution are 
\begin{align}
  \label{eq:joint-regular}
  p(s,t)= e^{-(s+t)},\quad p(r)=\frac{2}{(1+r)^2}.
\end{align}

\begin{figure}[h]
  \includegraphics[width=1\linewidth]{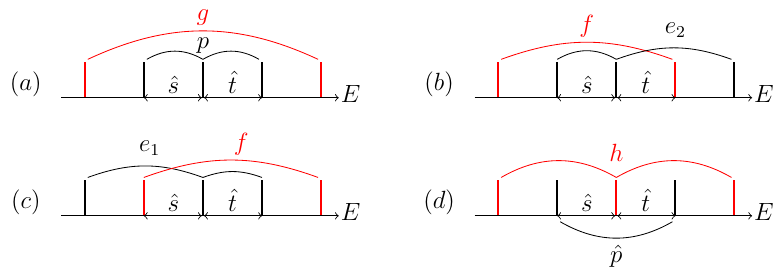}
  \caption{A scheme of the distribution of two consecutive spacings, where four configurations exist. Different colors denote spectra from different blocks.}
  \label{fig:sr-simple}
\end{figure}

\begin{figure*}
 \begin{centering}
   \includegraphics[width=0.85\linewidth,width=1.9\columnwidth]{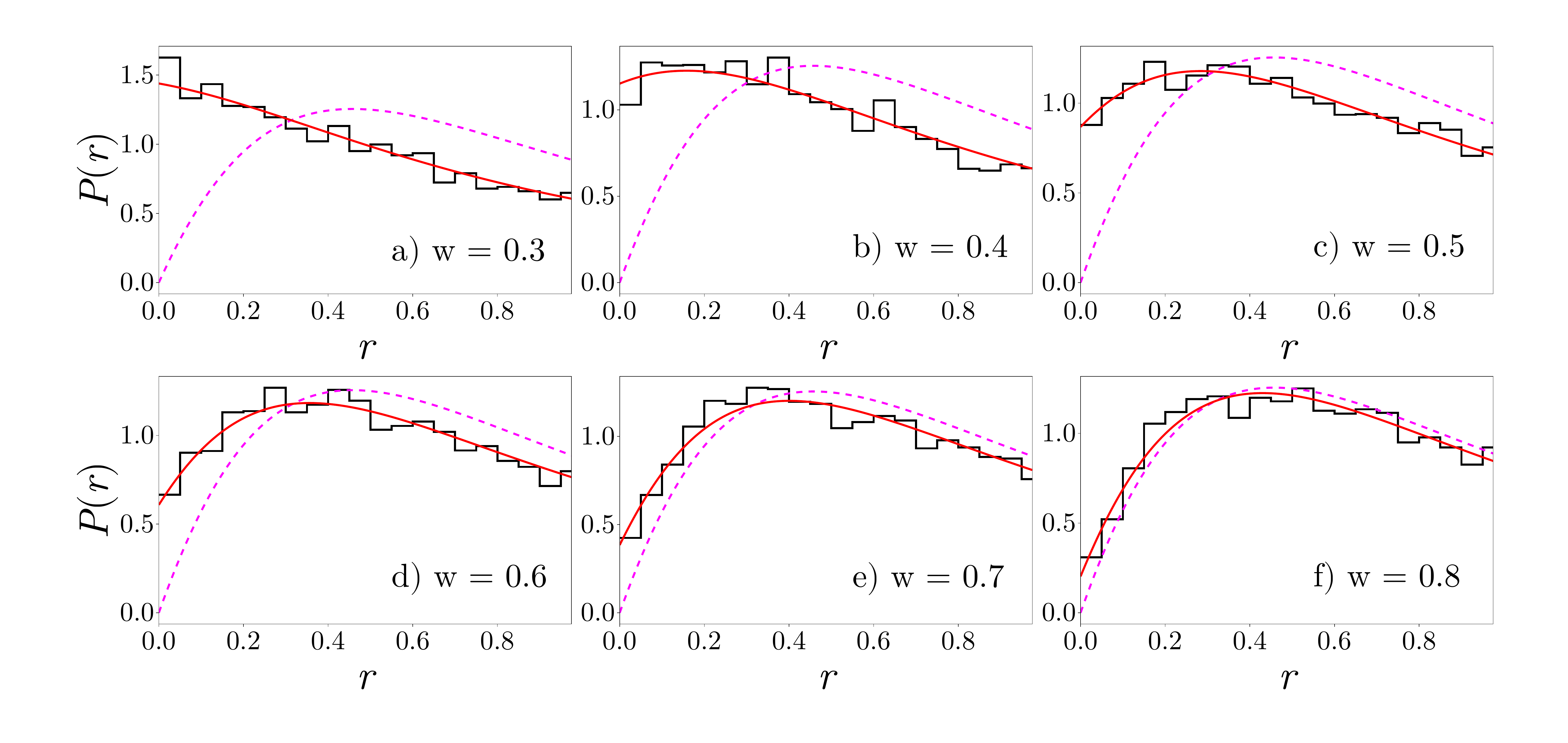}  
   \par\end{centering}
 \caption{Numerical level spacings ratio distribution compared with the analytical theory by Yan \cite{Yan2025}
   using the exact analytical
   value of $\mu_c$, $\rho (reg) = 1 - \mu_c$, for 6 geometries $w=0.3-0.8$ in equal steps of $0.1$,
    for $k\in \lbrack 970, 1000 \rbrack$
   from (a) to (f).  The black solid line represents the binned numerical data and the red solid line Yan's theory. In the background the GOE curve is shown for comparison.
   It is clearly seen that theoretical curve follows the numerics very well,
   showing that the localization effects are minor.}
\label{figPRallhigh}
\end{figure*}

\begin{figure*}
 \begin{centering}
   \includegraphics[width=0.85\linewidth,width=1.9\columnwidth]{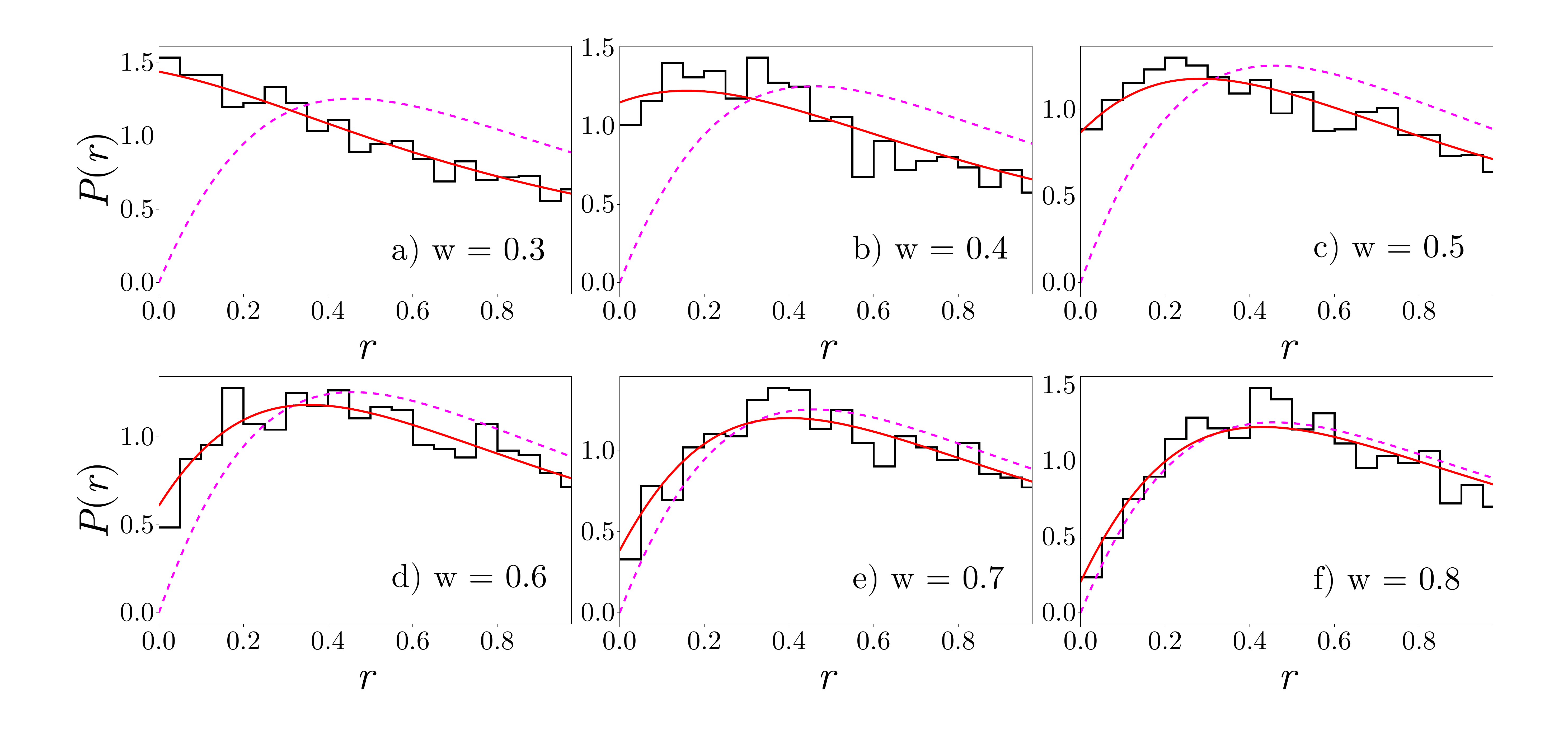}  
   \par\end{centering}
 \caption{The same as in Fig.\ref{figPRallhigh}, but for the spectrum $k\in \lbrack 5, 150 \rbrack$,
 showing that the effects of localization are minor, as the numerics and the theoretical curves agree very well.}
\label{figPRalllow}
\end{figure*}

Similarly to the gap probability approach used for the level spacing distribution, the spacing ratios distribution requires the introduction of \emph{two families of gap probability functions}. These are defined based on the joint distribution of consecutive spacings $p(s,t)$, as detailed in Ref. \cite{giraud2022probing}. Firstly, there are the one-variable functions:
  \begin{gather}
    \label{eq:gap-func1}
    f(s)=\int^\infty_0da\int^\infty_0db \ p(s+a,b), \nonumber \\
    g(s)=\int^\infty_0da \int^\infty_0db \int^\infty_0dc \ p(s+a+b,c),
  \end{gather}
and then the two-variable functions
  \begin{gather}
  \label{eq:gap-func2}
    e_1(s,t)=\int^\infty_0 da \ p(s+a,t), \nonumber\\
    e_2(s,t)=\int^\infty_0 da\int^\infty_0 db \ p(s,t+b),\nonumber\\
    h(s,t)=\int^\infty_0 da\int^\infty_0 db \ p(s+a,t+b).
  \end{gather}
These functions are related by the identities 
\begin{align}
  \label{eq:identities}
  \partial_s\partial_t h =p,\ \partial_s h =-e_1,\ \partial_t h=-e_2,\ \partial_s g=-f,
\end{align} 
 with 
 \begin{align}
  \partial_s f = -\hat{p}(s),\quad \hat{p}(s)=\int_0^\infty db\ p(s,b).
 \end{align}

Fig. \ref{fig:sr-simple} illustrates four different configurations for the distribution of two consecutive spacings, with the gap probability functions, where $s=\rho \hat{s}$, denotes the level spacing after unfolding, provided $\rho$ the density of states. The probability of each configuration can be expressed as a composition of various gap probability functions.
\begin{align}
  \text{(a)}\quad p_a(s,t) &= \mu_i^3 p(\rho_i \hat{s}, \rho_i \hat{t}) g(\rho_j(\hat{s}+\hat{t})) \notag \\
                           &= \mu_i [\partial_s \partial_t h(\mu_i s, \mu_i t)] g(\mu_j(s+t)), \notag \\
  \text{(b)}\quad p_b(s,t) &= \mu_i^2 \mu_j e_2(\rho_i \hat{s}, \rho_i \hat{t}) f(\rho_j(\hat{s}+\hat{t})) \notag \\
                           &= \mu_i [\partial_s h(\mu_i s, \mu_i t)] [\partial_t g(\mu_j(s+t))], \notag \\
  \text{(c)}\quad p_c(s,t) &= \mu_i^2 \mu_j e_1(\rho_i \hat{s}, \rho_i \hat{t}) f(\rho_j(\hat{s}+\hat{t})) \notag \\
                           &= \mu_i [\partial_t h(\mu_i s, \mu_i t)] [\partial_s g(\mu_j(s+t))], \notag \\
  \text{(d)}\quad p_d(s,t) &= \mu_i^2 \mu_j h(\rho_j \hat{s}, \rho_j \hat{t}) \hat{p}(\rho_i(\hat{s}+\hat{t})) \notag \\
                           &= \mu_j h(\mu_j s, \mu_j t) [\partial_s \partial_t g(\mu_i(s+t))]. \notag
\end{align}
Summing over $i\ne j$ for all configurations, we have
\begin{align}
  \label{eq:sr-simple}
  P(s,t)=\sum_{\mathcal{C}\in\{a,b,c,d\}}\sum_{i, j=1}^2 p_\mathcal{C}(s,t)=\partial_s\partial_t E(s,t), 
\end{align}
where 
\begin{align}
  E(s,t) =\sum_{i=1}^2 \mu_ih(\mu_is,\mu_i t)\prod_{j\ne i}g(\mu_j(s+t)).
\end{align}
The resulting distribution of spacing ratios can be derived from \( P(s, t) \), as shown in Eq.~\eqref{PrPoissonWigner} as 
\begin{align}
    P(r)=2\int_0^\infty  P(s,rs)sds. 
\end{align}
It cannot be given in a closed formula, as it rests upon quite complicate integrals of special functions, but nevertheless
gives an analytical, theoretical, result for any value of the Berry-Robnik parameter $\mu_c$,
on the level of Wigner-type approximation, (\ref{PSWigner}) and (\ref{PrPoissonWigner}), for $2\times 2$ and
$3 \times 3$ GOE matrices, respectively. It is thus
tempting to explore the $P(r)$ for the energy spectra of our 6 geometries of the mushroom billiards. In Fig. \ref{figPRallhigh} we show $P(r)$ for the high-lying states. The agreement is perfect.
In Fig. \ref{figPRalllow} we plot $P(r)$ for low-lying states, showing that also here the
localization effects are quite weak.

%

\begin{figure}
 \begin{centering}
   \includegraphics[angle=0,width=0.95\linewidth]{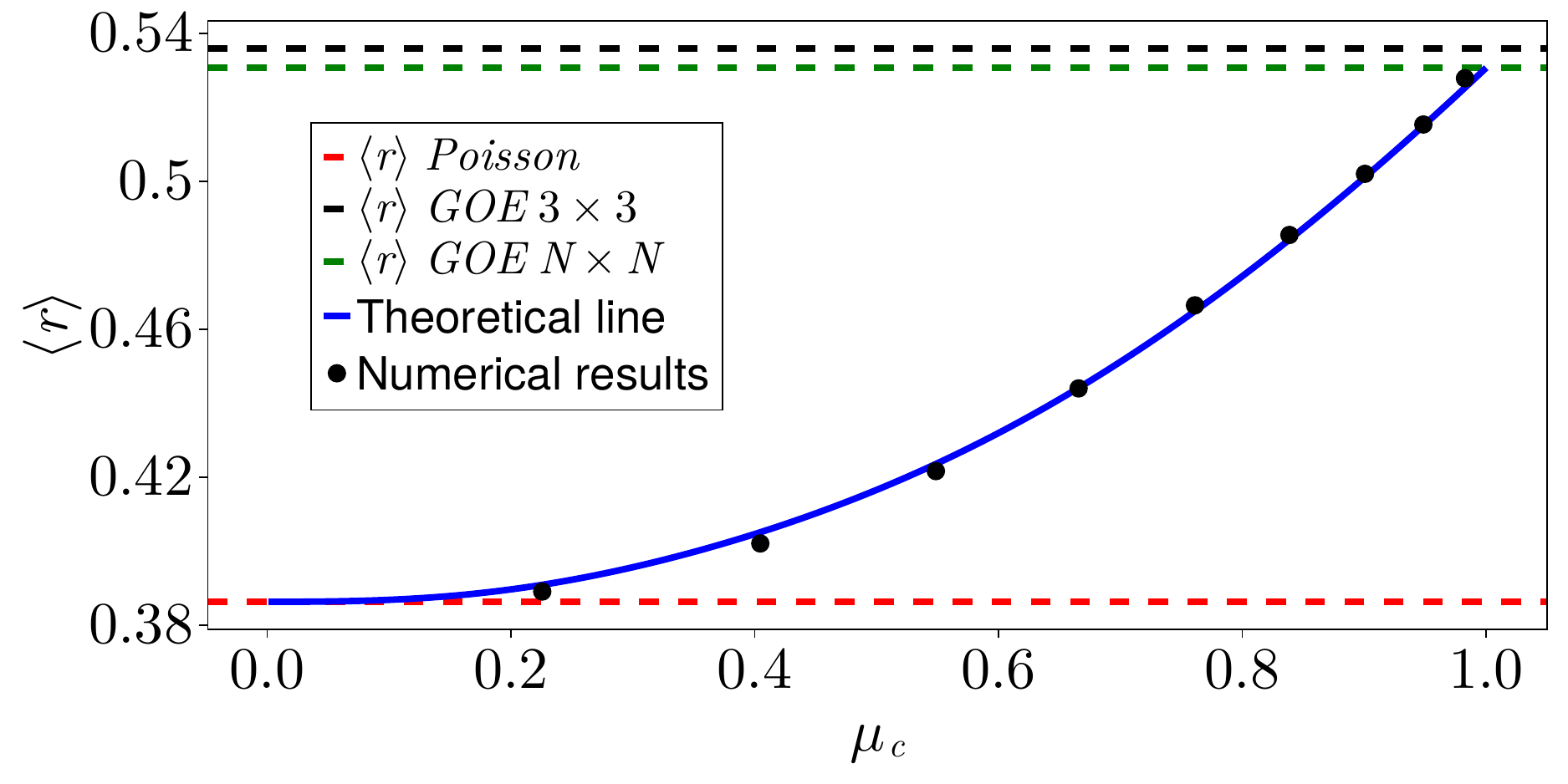}  
   \par\end{centering}
 \caption{The mean level spacing ratio compared with the analytical theory by Yan \cite{Yan2025} $GOE\;N\times N$,
   using the exact analytical
   value of $\mu_c$, $\rho (reg) = 1 - \mu_c$, for 9 geometries $w=0.1-0.9$ in equal steps of $0.1$,
    for $k\in \lbrack 900, 1000 \rbrack$.
   It is clearly seen that theoretical curve follows the numerics very well,
   showing that the localization effects are minor, except for $w=0.1$ and $0.2$.}
\label{figmrchigh}
\end{figure}

It is practical to characterize the $P(r)$ by its mean value,
$\langle r \rangle = \int_0^1 r P(r)dr$. For the Poisson case we have
$\langle r \rangle_P = 2\ln2-1 \approx 0.3863$ and
for the Wigner case $\langle r \rangle_W =4-2\sqrt{3} \approx 0.5359$ \cite{AtaBogGirRou2013}. 
For the mixed case with $\mu_c$ between $0$ and $1$ the
theoretical result by Yan accurately predicts the numerical value as shown in Fig. \ref{figmrchigh}.
It corresponds to the GOE result  for asymptotics where the dimension of the matrices $\rightarrow \infty$,
after a rescaling as explained in Ref. \cite{Yan2025}.

%

%
The conclusion of the Sections \ref{sec4} and \ref{sec5} is that we have almost perfect agreement between
the Berry-Robnik theory \cite{BerRob1984} for the level spacings distribution $P(s)$ for the geometries
$w=0.3 - 0.9$, as well the agreement between Yan's theory \cite{Yan2025} for spacings ratio distribution $P(r)$
in the mixed-type regime, 
showing that the localization effects are indeed very weak. They will be studied in detail in paper II
\cite{OreLozRobYan2025}, by the analysis of Poincar\'e-Husimi functions of the eigenstates,
especially in the geometries $w=0.1$ and $0.2$, where the classical dynamics is quite complex and needs
an extended analysis.

\section{The effects of localization as reflected in the energy spectra}
\label{sec6}

\begin{figure*}
 \begin{centering}
   \includegraphics[width=0.85\linewidth]{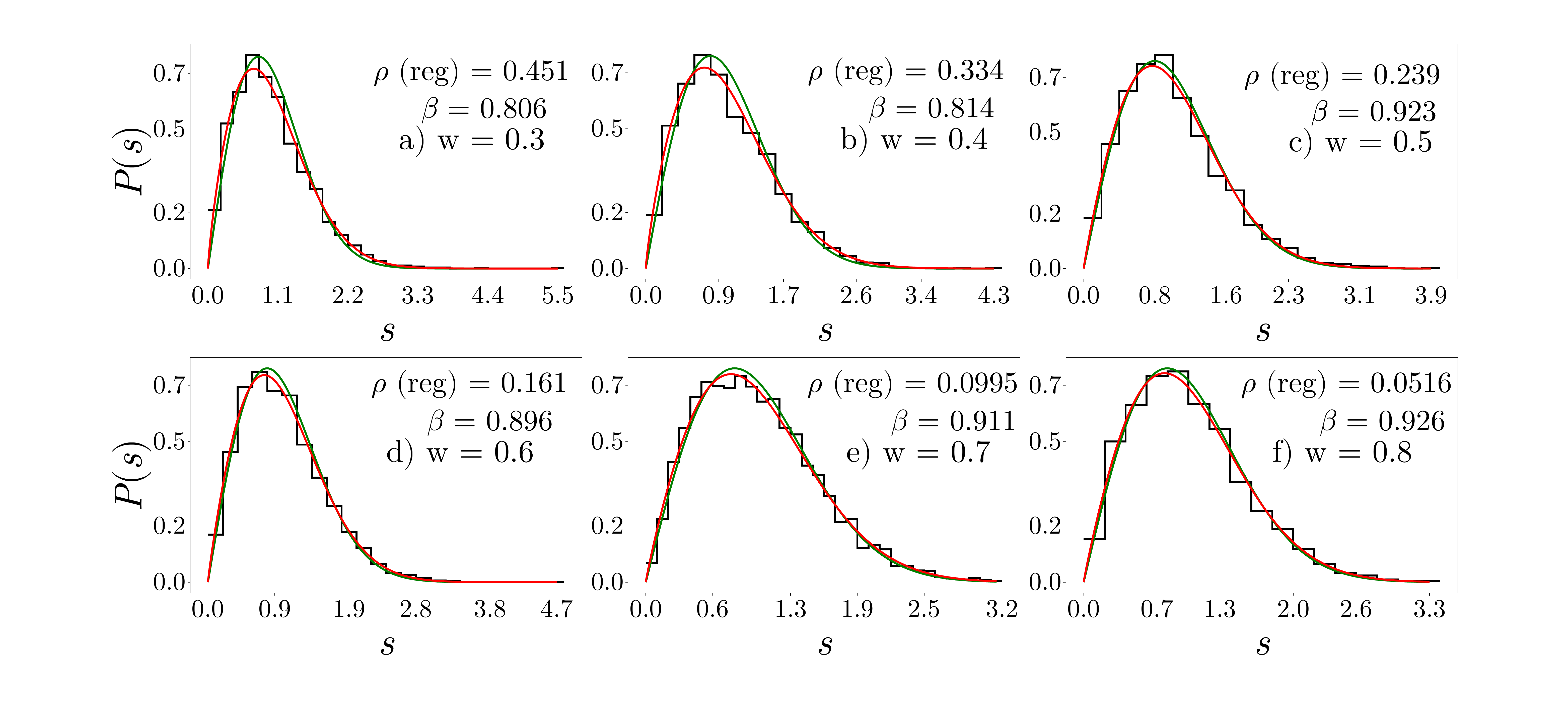}  
   \par\end{centering}
 \caption{Numerical level spacings distribution for the chaotic
   states in the interval $k \in \lbrack 970, 1000 \rbrack $
   compared with the Brody distribution (\ref{PSB})
   after the separation using the classical criterion,
   for 6 geometries $w=0.3-0.8$ in equal steps of $0.1$,
   from (a) to (f). The green solid line represents the BR distribution and the red solid one the BRB distribution. The
   difference demonstrates weak localization effects, so that
   $\beta$ is almost equal to $1$, considering the statistical
   error of the order $\approx \pm 0.1$.}
\label{figPSchaotichigh}
\end{figure*}

\begin{figure*}
 \begin{centering}
   \includegraphics[width=0.85\linewidth]{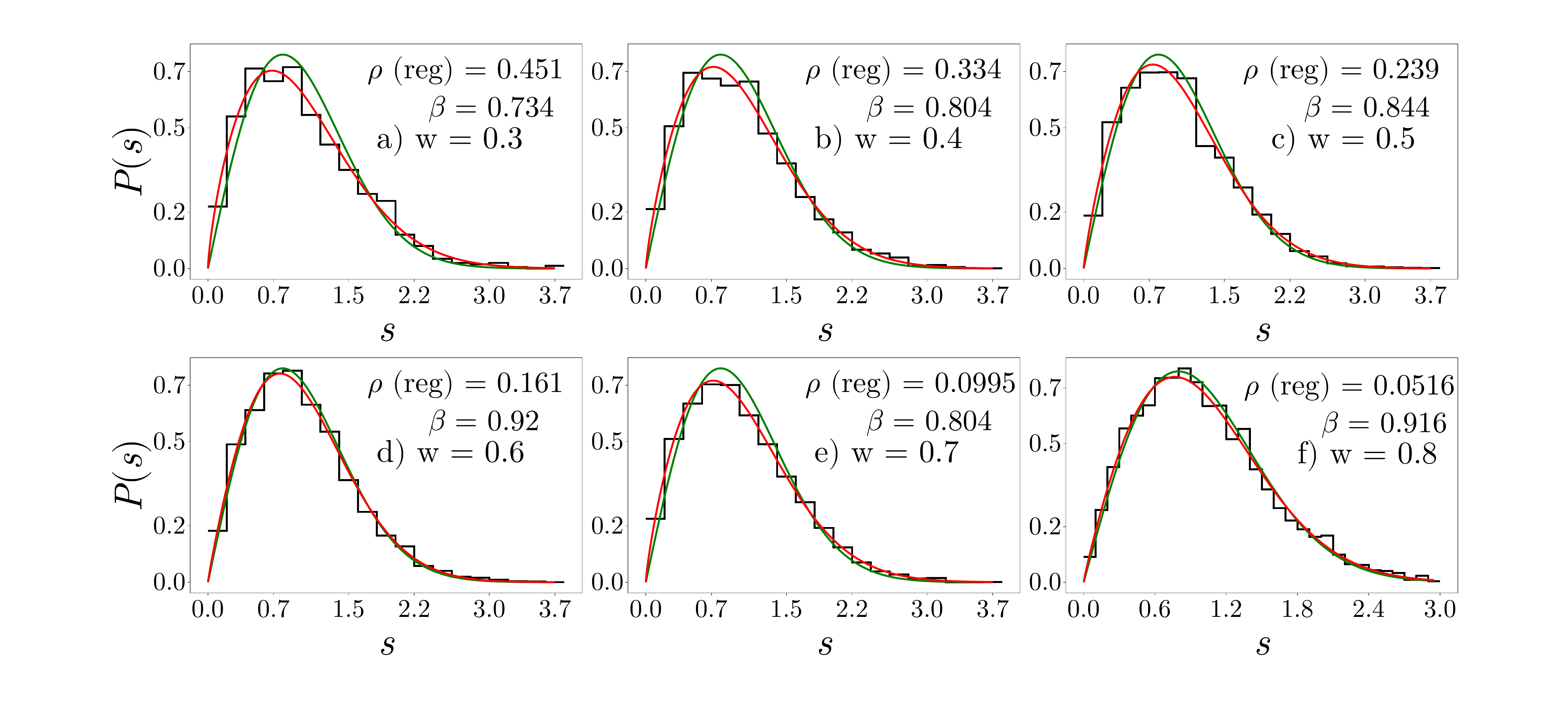}  
   \par\end{centering}
 \caption{The same as in Fig.\ref{figPSchaotichigh}, but for $k\in \lbrack 5, 150\rbrack$,
   showing that even at low energies $k^2$ the effects of localization also are minor,
   but definitely larger than at high-lying states $k \in \lbrack 970, 1000 \rbrack $
   of Fig. \ref{figPSchaotichigh}.}
\label{figPSchaoticlow}
\end{figure*}

So far it has been demonstrated that the localization effects as reflected in the spectral
statistics are quite weak, except possibly for $w=0.1$ and $w=0.2$. It must be emphasized
that the determination of the Brody level repulsion parameter  $\beta$ by the best fitting
BRB distribution is very badly defined, as the minimum of the underlying distance between the
numerical data and the best fitting BRB is quite shalow, implying that the estimated error
in the value of $\beta$ can be of the order $\pm 0.2$.

Nevertheless, in order to see more clearly the effects of localization, we now employ the
method of separating the regular and the chaotic eigenstates, as introduced by Batistić and
Robnik \cite{BatRob2013A,BatRob2013B}, based on the structure of Poincar\'e-Husimi functions as
compared with the classical phase space portrait (see details in Appendix. \ref{appB}). This approach will be extensively applied
in paper II \cite{OreLozRobYan2025}. Here we just note that we have separated the regular and
chaotic states based on the classical criterion of dividing the states between regular and chaotic in
the proportion of classical $\mu_r=1-\mu_c$ and $\mu_c$.

Then, the spectral statistics $P(s)$ has been calculated for the chaotic part.
By the best fitting procedure using equation (\ref{PSB}) the Brody exponent $\beta$ has been
determined. In Fig.\ref{figPSchaotichigh} we show $P(s)$ for the high-lying states, where
the weak localization effects are detected. In order to demonstrate that the localization effects are minor also at low-lying states, we show the
similar plot in Fig.\ref{figPSchaoticlow}. The observation of this trend is finally corroborated by the tendency of
$P(r)$ distribution towards
GOE at high energies as shown in Fig.\ref{figPRchaotichigh}. The effects of localization definitely become larger at smaller $k$,
as shown in Fig.\ref{figPRchaoticlow}, but are nevertheless small.

In conclusion of this section we can maintain that the localization effects
are present, but are weak, and they are stronger at lower energies $k^2$, as expected
theoretically, or in other words, we see that the localization effects
gradually disappear in the semiclassical limit $k \rightarrow \infty$,
in accordance with the Principle of Uniform Semiclasisical Condensation (PUSC)
of Wigner or Husimi functions \cite{Rob1998,Rob2023Rev}.

\section{Discussion, conclusions and outlook}
\label{sec7}

In this paper I we have extensively studied the spectral statistics of a large family of
mushroom billiards introduced by Bunimovich, which are mixed-type systems, but not of the KAM type.
Instead, each billiard of this family has exactly one regular and one chaotic (ergodic)
component with a smooth boundary between them, which can be analytically treated, so that
we have exact theoretical closed formula for the Berry-Robnik parameter $\mu_c$ \cite{BerRob1984}.
We have analyzed geometries with the stem width $w=0.3 - 0.9$, where the system behaves
very well in agreement with the theory, while $w=0.1$ and $w=0.2$ will be treated separately
due to the more complex structure of the classical phase space and consequently some
intricate and intriguing quantum effects.

Therefore, we can compare the theoretical level spacings distribution with the numerical
histogram, showing excellent agreement at high-lying eigenstates, while at lower energies
we see effects of weak localization, which are well captured by the Berry-Robnik-Brody distribution.
After separation of regular and chaotic eigenstates using the method of Batistić and Robnik,
presented in detail in paper II, we find that the chaotic eigenstates indeed obey the
Brody distribution, although the level spacings distribution is close to GOE.

We also have studied the level spacings ratio distribution, and compared it with the
recent theory by Yan \cite{Yan2025}. At high energies the agreement is perfect, as the
localization effecs are weak, but at lower energies one can see small deviations of
the histogram from the theoretical curve, just due to the weak localization effects.

It is well understood that the localization effects are weak at large $k$, expected on
the basis of Heisenberg time scale in comparison with the classical
transport time,  giving values of the parameter $\alpha=1.77\; k/N_T >> 1$. 
At smaller $k$, the localization is stronger, although
still being weak. We may conclude that the statistical properties of the energy spectra in this mixed-type
regime are theoreticaly well understood in terms of Berry-Robnik picture for the
level spacings distribution and in terms of Yan's theory for the level spacings
ratio distribution. The picture is quite different for geometries $w=0.1$ and $w=0.2$, as the
classical dynamics is more complex, which will be treated
separately in another work.

In paper II we shall perform a complete calculation of the Poincar\'e-Husimi functions
for a large range of eigenstates at various energies, for all geometries $w=0.1-0.9$,
and consider the localization properties, the criterion and the method of separation
of regular, mixed and chaotic eigenstates. We also shall show that the relative fraction
of mixed-type eigenstates decays as a power law in the semiclassical limit, as a function
of a semiclassical parameter (effective Planck constant).

\section{Acknowledgement}

This work was supported by the Slovenian Research and Innovation
Agency (ARIS) under the grants J1-4387 and P1-0306.

\appendix

\begin{figure*}
 \begin{centering}
   \includegraphics[width=0.8\linewidth]{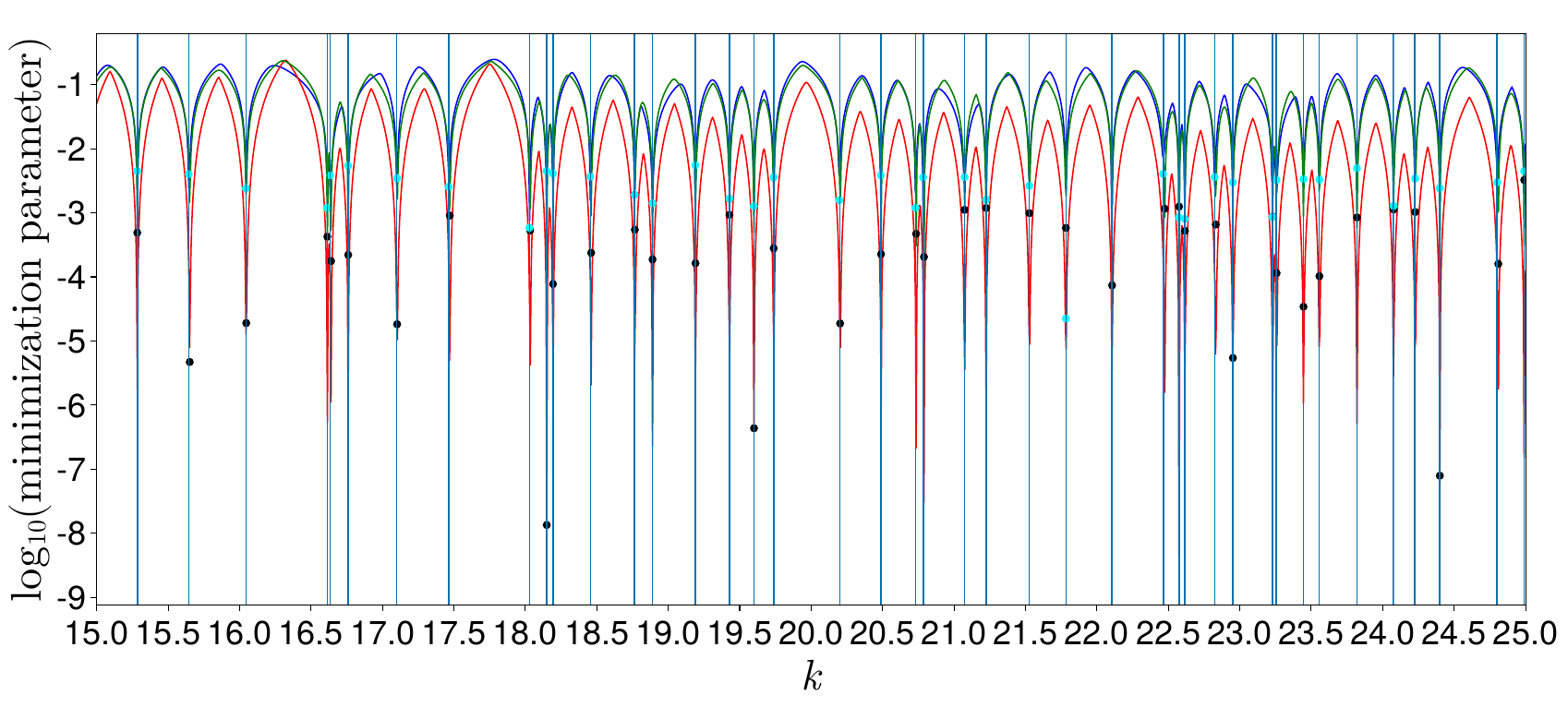}  
   \par\end{centering}
 \caption{Comparison of minimization parameter plots on the interval $15\leq k  \leq 25$ for different methods:
   Boundary integral method (blue),
   Heller's plane wave decomposition method (green),
   particular solutions method - extended Heller's method (red), Vergini-Saraceno
   method - Fourier-Besel version (black dot, representing the minimum of tension),
   expanded boundary integral method (Veble et al \cite{VebProRob2007}, light blue dot).
   The blue vertical line represents the eigenvalue of all methods.}
\label{figmethods1}
\end{figure*}

\section{Numerical techniques}
\label{appA}

We briefly describe different boundary-integral and basis-expansion methods that were used for the comparison of quantum eigenvalues in the (non-analytic) mushroom billiard.  Because the boundary is not analytic, all methods not using the corner-adapted Fourier-Bessel basis exhibit non-exponential convergence regardless of quadrature choice~\cite{kress,betcke}.  Our primary goal is to test the Corner-Adapted Fourier-Bessel basis (which avoids the convergence issues of plain plane-wave or Bessel bases in non-convex billiards with corners~\cite{Gutkin2003,Barnett2000}).  We organize the methods into two families—those that discretize the boundary integral with Hankel‐kernel quadrature, and those that use Bessel‐kernel expansions—always sampling the collocation points on the boundary. The quadrature used was the trapezoidal rule for methods that used Hankel functions and Gauss-Legendre otherwise.
A minimization parameter plot for comparing each of these methods is given in Fig. \ref{figmethods1} showing convergence of different methods to the same eigenvalue (given by the blue vertical line).
 
\subsection{Using Hankel functions}

\subsubsection{Boundary Integral Method (BIM)}
\label{nm1}
We solve for the smallest singular value of the discretized double-layer potential boundary operator which is the Fredholm matrix $A$. We first start with constructing the integration kernel
\begin{align}
    Q_{k}(s,s') = -\frac{ik}{2} \cos \phi(s,s') H_{1}^{(1)} (k\tau (s,s')),
\end{align}
where $s,s'$ are arclength coordinates on the boundary with $s \neq s'$, $q=(x,y)$, $ q'=(x',y')$ are the coordinates on the boundary, $\tau$ is the Euclidian distance between points on the boundary $\tau = \sqrt{(x-x')^2+(y-y')^2}$ and $k$ is the wavenumber parameter. The normal vector $n(s)$ is found in the term
\begin{align}
    \cos \phi (s,s') = \frac{n(s) \cdot (q(s) - q(s'))}{\tau (s,s')}.
\end{align}
$ Q_{k}(s,s')$ needs to be symmetrized in such a way that we get it's value to be $0$ on the mushroom billiard symmetry line for the odd-odd symmetry class. We also use the limit $Q(s,s' \rightarrow s) = \frac{1}{2\pi} \kappa(s)$ where $\kappa(s)$ is the curvature at arclength $s$ to handle divergences. From this we construct the Fredholm matrix (where we assume constant difference between neighbouring arclengths $\Delta s$)

\begin{align}
    A_{k}(s(i),s'(j)) = \delta_{ij} - \Delta s Q_{k}(s(i),s'(j)).
\end{align}

The true eigenvalues are obtained as the smallest singular values of the Fredholm matrix \(A(k)\). However, assembling \(A(k)\) and computing its SVD at each trial \(k\) is prohibitively expensive when many eigenvalues are needed.  In the next subsection we therefore introduce an accelerated scheme that reuses the same boundary operator to compute potentially many eigenvalues per solve, yielding a substantial speedup.  Finally, in non-convex geometries one must also include the first-layer boundary operator to eliminate exterior solutions~\cite{kress}.

\subsubsection{Expanded Boundary Integral Method (EBIM)}
\label{nm2}
This method is an improvement of BIM because we are able to potentially get many eigenvalues in a small interval $[k_{0} - \delta k, k_{0} + \delta k]$ around a trial eigenvalue $k_{0}$. Taylor expansion of the BIM condition $A(k)\mathbf{u}=0$ gives

\begin{align}
    A(k_{0}+\delta k)\mathbf{u} = [A(k_{0}) + \delta k \frac{\partial A(k_{0})}{\partial k} + O(\delta k^2)]\mathbf{u}=0.
\end{align}

\noindent Following the original paper \cite{VebProRob2007} we obtain the following generalized eigenvalue problem (GEP)

\begin{align} \label{eq:gep_EBIM}
    A(k_{0})\mathbf{u} = \lambda \frac{\partial A(k_{0})}{\partial k} \mathbf{u},\quad  \mathbf{v}^{\dagger} A(k_{0}) = \lambda \mathbf{v}^{\dagger} \frac{\partial A(k_{0})}{\partial k},
\end{align}

\noindent where the matrices $A(k)$ and $\frac{\partial A(k)}{\partial k}$ are evaluated at the trial eigenvalue $k_{0}$. We solve it for the smallest $\lambda$s, specifically those that satisfy $|\lambda|<\delta k$. It must be noted that both $A$ and $\frac{\partial A}{\partial k}$ are non-Hermitian matrices and therefore their GEP gives complex generalized eigenvalues. We therefore impose an additional criterion that $\Im(\lambda)<10^{-2}\Re(\lambda)$. Similary to BIM we construct the Fredholm matrix as described in the BIM section and formulate the $1^{st}$ and $2^{nd}$ $k$-derivative of the integration kernel $Q_{k}(s,s')$

\begin{align}
&\frac{\partial A_{k}(s,s')}{\partial k}
  = -\,\Delta s \frac{\partial Q_{k}(s,s')}{\partial k} \notag \\[6pt]
  &= \Delta s\frac{i k}{2}\cos\phi(s,s')\,\tau(s,s')\,
     H_{0}^{(1)}\bigl(k\,\tau(s,s')\bigr) ,
\end{align}
and 
\begin{multline}
\frac{\partial^2 A_{k}(s,s')}{\partial k^2}
= -\,\Delta s\;\frac{\partial^2 Q_{k}(s,s')}{\partial k^2}
\\
\shoveleft
= -\,\Delta s\,\frac{i}{2k}\,\cos\!\phi(s,s')\,
\Bigl[\bigl(-2 + (k\,\tau(s,s'))^2\bigr)\,
\\
\times H_{1}^{(1)}\bigl(k\,\tau(s,s')\bigr) + k\,\tau(s,s')\,H_{2}^{(1)}\bigl(k\,\tau(s,s')\bigr)\Bigr],
\end{multline}

\noindent with the symbols defined in the BIM section. The diagonal limits are $\frac{\partial Q_{k}(s,s' \rightarrow s)}{\partial k} = \frac{\partial^2 Q_{k}(s,s' \rightarrow s)}{\partial^2 k} = 0$. Similarly to BIM we symmetrize the $k$-derivatives of $Q_{k}(s,s')$ such that their values are $0$ on the symmetry line of the mushroom billiard geometry (odd symmetry class). This allows us to write the potential eigenvalue solution in the given interval as

\begin{align}
    k = k_{0} - \lambda - \frac{\lambda^2}{2} \left( \mathbf{v}^\dagger \frac{\partial^2 A(k_{0})}{\partial^2 k} \mathbf{u} \right) \left(\mathbf{v}^\dagger \frac{\partial A(k_{0})}{\partial k}\mathbf{u} \right)^{-1}.
\end{align}

\noindent It is important to note that while this method becomes more efficient at higher trial eigenvalues $k_{0}$ its condition number also rises (among others due to Weyl's law as the eigenvalues become more densely spaced and we pick our $k_{0}$ closer to the true eigenvalue for which the $\det A(k_{0}) = 0$). Eq. (\ref{eq:gep_EBIM}) is typically ill-conditioned with the condition number of the relevant matrices being $\text{cond} (A(k_{0})) \approx 10^{3}$ and $\text{cond}(\frac{\partial A(k_{0})}{\partial k}) > 10^{12}$ in the neighborhood of a true eigenvalue. How to properly regularize/reformulate this GEP is an open problem.

\begin{figure}
    \centering
    \includegraphics[width=8cm]{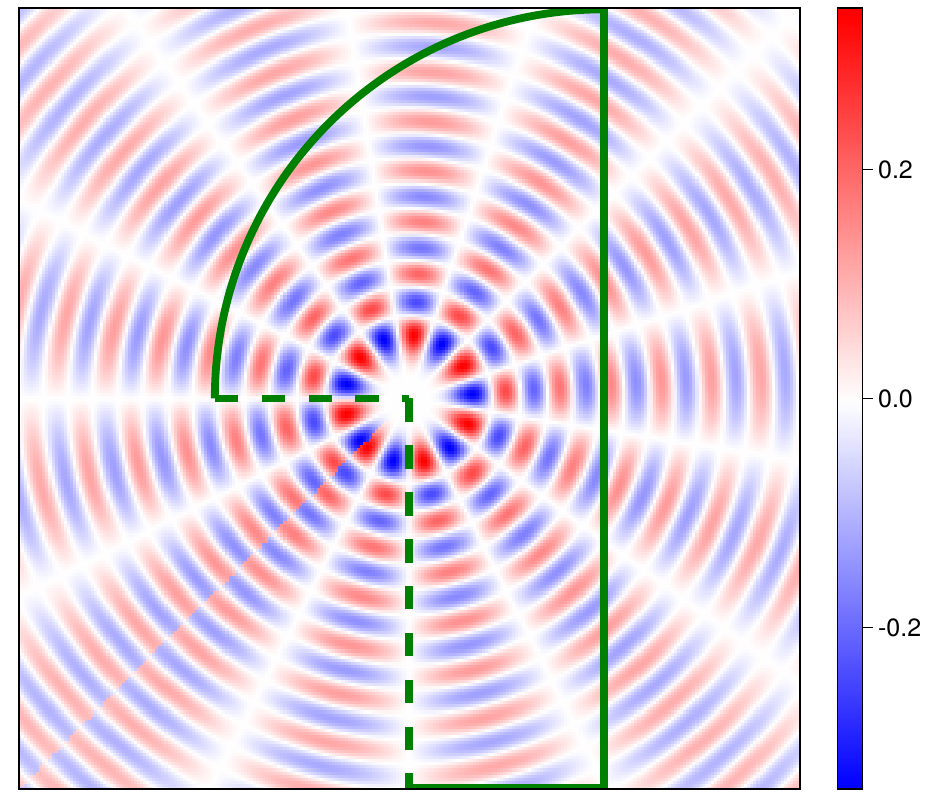}
    \caption{Example of a Fourier-Bessel basis function with a rotated discontinuity showing automatically satisfied boundary conditions on the segments forming the reentrant corner}
    \label{basis}
\end{figure}

\subsection{Using Corner-Adapted Fourier-Bessel functions}

In our approach we expand each eigenfunction as a finite linear combination of corner–adapted Fourier–Bessel modes:
\begin{align}
   \Psi_{k}(r,\phi)\;\approx\;\sum_{i=1}^{N} x_{i}\,f_{i}(k,r,\phi). 
\end{align}
In this study, we choose the basis functions
\begin{align}
  f_{i}(k,r,\phi)
=J_{\tfrac{2i}{3}}(k\,r)\,\sin\!\bigl(\tfrac{2i}{3}\,\phi\bigr),  
\end{align}
which incorporate the correct reentrant–corner behavior.

\noindent The basis function is constructed from Bessel functions of the first kind $J_{i}$ of order $i$ which hold the $r$ dependence and an angular part for the $\phi$ dependence (where $(r,\phi)$ are the polar representations of the point $(x,y)$ wrt. the reentrant corner as the origin). In Fig.\ref{basis} we plot this basis with the reentrant corner $\frac{3\pi}{2}$ superimposed with the mushroom billiard boundary. One also requires the $\frac{\partial f}{\partial x}$ and $\frac{\partial f}{\partial y}$ of the basis functions from which the normal derivative of the basis functions is constructed

\begin{align}
    \frac{\partial f_{i}(k,r,\phi)}{\partial x} &= \frac{kx}{r} J'_{\frac{2}{3}i}(k r) \sin \frac{2i }{3}\phi  - \frac{2iy}{3r^2}  J_{\frac{2}{3}i}(k r) \cos \frac{2i }{3}\phi,\nonumber \\
    \frac{\partial f_{i}(k,r,\phi)}{\partial y} &= \frac{ky}{r} J'_{\frac{2}{3}i}(k r) \sin \frac{2i }{3}\phi  + \frac{2ix}{3r^2} J_{\frac{2}{3}i}(k r) \cos \frac{2i }{3}\phi, \nonumber \\
    \frac{\partial f_{i}(k,r,\phi)}{\partial n} &= n_{x} \frac{\partial f_{i}(k,r,\phi)}{\partial x} + n_{y} \frac{\partial f_{i}(k,r,\phi)}{\partial y}.
\end{align}

\subsubsection{Decomposition Method (DM)}
\label{nm3}
The idea behind this method is to minimize the boundary tension $F(k)$ with a general boundary weight function $w(s)$ (we choose $w(s)=1$) defined as

\begin{align}
F(k) = \oint ds \;w(s)\,\bigl|\Psi\bigl(r(s),\phi(s)\bigr)\bigr|^2 
  = \sum_{i,j=1}^{N} x_{i}\,F_{ij}(k)\,x_{j},
\end{align}
with $F_{ij}(k) = \oint ds w(s) f_{i} (k,r(s),\phi (s)) f_{j} (k,r(s),\phi (s))$ being the basis boundary tension matrix. We also define the usual normalization condition as follows:

\begin{align}
1 = \iint_{D} dS \,\bigl|\Psi_{k}(r,\phi)\bigr|^2 
  = \sum_{i,j=1}^{N} x_{i}\,G_{ij}(k)\,x_{j},
\end{align}
with $G_{ij}(k) = \iint_{D} dS f_{i}(k,r,\phi) f_{j} (k,r,\phi)$ and  $\iint_{D}$ being the area integral in the interior of the boundary. This calculation would be very expensive as it would require the evaluation of an area integral and is never done in practice. Rather an equivalent version of this integral is performed that is valid only when the wavefunction vanishes on the boundary~\cite{Barnett2000} (non-applicable to Neumann BCs on the symmetry lines)

\begin{multline}
    G_{ij}(k) = \oint ds \left(\frac{r_{n}(s)}{2k^2} \right) \left(\frac{\partial f_{i}(k,r(s),\phi (s))}{\partial n}\right) 
    \\
    \times \left(\frac{\partial f_{j}(k,r(s),\phi (s))}{\partial n}\right).
\end{multline}

\noindent With these $F_{ij}(k)$ and $G_{ij}(k)$ we formulate the problem of finding eigenvalues and eigenvectors of the original problem $(\Delta + k^2) \Psi = 0$ with $\Psi|_{\partial\Omega}$ as a minimization condition with an additional constraint coming from the normalization condition

\begin{align}
    \min \left(\mathbf{x}^T F \mathbf{x} \right) \quad \text{with} \quad \mathbf{x}^T G \mathbf{x} = 1.
\end{align}

\noindent This can be rewritten as an equivalent GEP that is easier to numerically handle

\begin{align}
    F(k) \mathbf{u} = \lambda G(k) \mathbf{u},
\end{align}

\noindent where the true eigenvalues are found at values of $k$ where $\lambda$ is a local maximum.

\subsubsection{Particular Solutions Method (PSM)}
\label{nm4}
This method is an extension of the original Heller's method where we approximate the 2d integral of $G_{ij}(k)$ described above with a sum over a large number of interior points $M$ 

\begin{align}
    G_{ij}(k)
    =\frac{A}{\lvert M\rvert}
    \sum_{(r,\phi)\in M}
       f_{i}(k,r,\phi)\,f_{j}(k,r, \phi)\,,
\end{align}
where $A$ is the area of the billiard and the basis functions $f_i$ as defined above. Everything else stays the same and we solve the same GEP as in the Decomposition method.

\subsubsection{Vergini-Saraceno ``Scaling" Method (VS)}
\label{nm5}
The fundamental idea behind this method is to use an expansion of the boundary tension $F$ at a trial eigenvalue $k_{0}$ up to first order and forming a GEP that can give us potentially many eigenvalues in a very small interval ($\delta k \approx 0.01k^{-1/3} $). We use a special weight function $w(s) = \frac{1}{r_{n}(s)}$ where $r_{n}(s) = n(s) \cdot r$ is the dot product of the normal vector at the boundary evaluated at $r$. The choice of this weight is crucial since the basis tension matrices are quasi-orthogonal with this weight~\cite{Barnett2000}. Expansion of $F$ at a trial eigenvalue $k_{0}$ gives

\begin{align}
    F(k_{0}) + \delta k \frac{\partial F(k_{0})}{\partial k} + O(\delta k^2) = 0.
\end{align}

\noindent For the tension matrix $F$ we the "special" weight function

\begin{align}
    F(k) = \oint \frac{ds}{r_n(s)} |\Psi_k (r(s),\phi(s))|^2.
\end{align}

\noindent This matrix can be rewritten as a basis tension matrix $\hat{F}(k)$

\begin{align}
    F(k) = \mathbf{x}^T \hat{F}(k) \mathbf{x} \quad \text{and} \quad \frac{\partial F}{\partial k} = \mathbf{x}^T \frac{\partial \hat{F}}{\partial k} \mathbf{x},
\end{align}
with the basis tension matrix element given by 
\begin{align}
    \hat{F_{ij}}(k) = \oint \frac{ds}{r_n(s)} f_{i} (k,r(s),\phi (s)) f_{j} (k,r(s),\phi (s)).
\end{align}
With these matrices we construct the GEP in the interval $[k_{0}-\delta k,k_{0}+\delta k]$
\begin{align} \label{eq:gep_VS}
    \hat{F}(k_{0}) \mathbf{x} = -\lambda \frac{\partial \hat{F}(k_{0})}{\partial k} \mathbf{x},
\end{align}
and look for those generalized eigenvalues that satisfy $|\frac{2}{\lambda} + O(\lambda ^2)| < \delta k$ since for those we are guaranteed to have true eigenvalues at $k = k_{0} - \frac{2}{\lambda} + O(\lambda ^2)$. The eigenvectors $x$ are exactly the basis expansion coefficients of the wavefunction in the chosen basis. In practice solving Eq. (\ref{eq:gep_VS}) is unstable since both $F(k)$ and $\frac{\partial F(k)}{\partial k}$ become numerically singular at the necessary (semiclassical) number of basis functions and collocation points \cite{Barnett2000}. Therefore we diagonalize $F$ and remove the eigenvalues whose absolute value is close to machine precision (we used $\epsilon = 10^{-15}$). These eigenvalues would give basis coefficients $x$ whose wavefunction inside the billiard gives near-zero contribution and are classified as spurious. This procedure removes the numerical nullspace of $F$ and makes inverting $\Lambda$ below numerically stable. Therefore we can transform the GEP to a regular eigenvalue problem by diagonalizing \( F(k_0) \) (real symmetric) as

\[
F(k_0) = S \Lambda S^\top,
\]
and retain only the eigenvalues \( \lambda_i \) such that \( \lambda_i > \epsilon \max(\Lambda) \). Let \( C \) be the matrix of retained eigenvectors, and define the scaled basis as
\[
\widetilde{C} = C \Lambda^{-1/2}.
\]
We project the generalized eigenproblem into the reduced basis
\[
\widetilde{C}^\top\, \frac{\partial F(k_0)}{\partial k}\, \widetilde{C} \, \mathbf{z} = \mu\, \mathbf{z},
\]
and after solving for the eigenpairs \( (\mu, \mathbf{z}) \), we recover the physical solutions of the original problem as
\[
\mathbf{x} = \widetilde{C} \mathbf{z} \quad \text{and} \quad \lambda_{\text{non-spurious}} = \mu.
\]

\begin{figure}
    \centering
    \includegraphics[width=8cm]{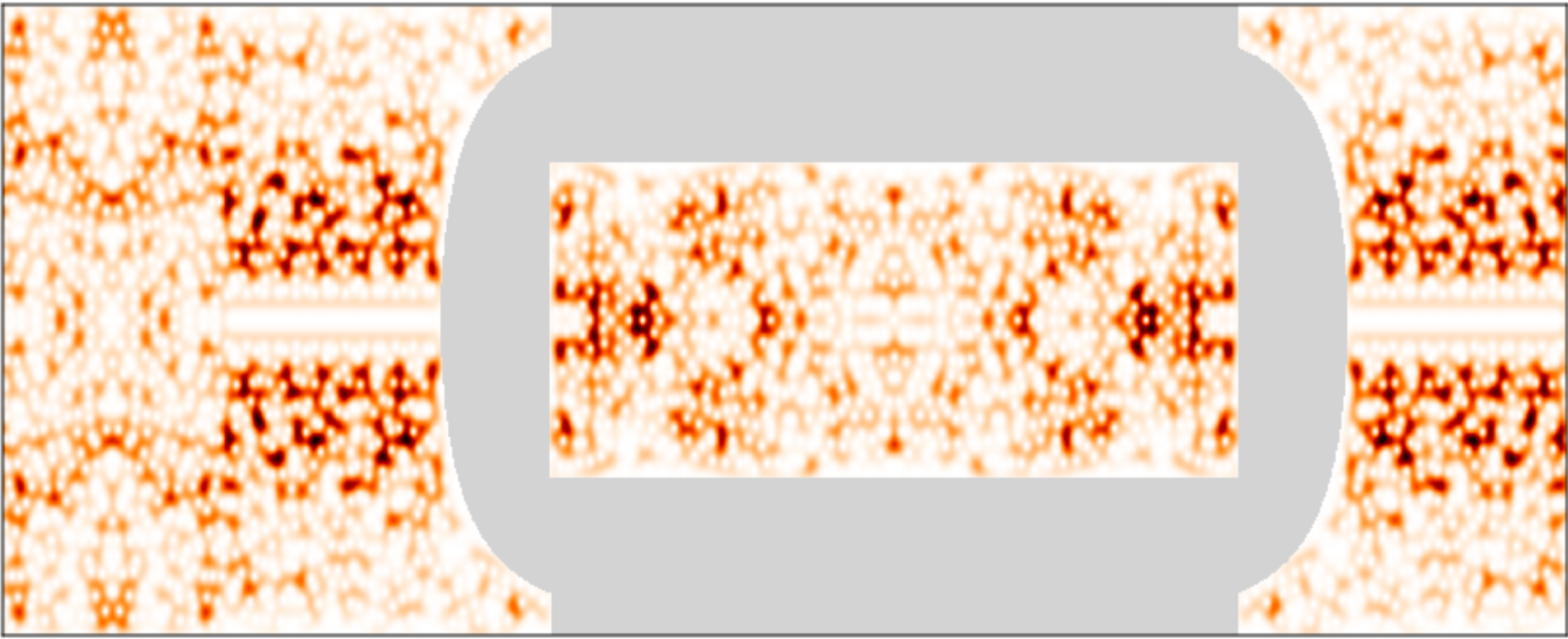}
    \caption{Example of a well spread chaotic Poincaré - Husimi function with the overlap grid of the regular region as gray. It was taken deep in the semiclassical limit so it's $M$ index value is $1$ due to  having no overlap with the classical regular region.}
    \label{figHusimi}
\end{figure}

\begin{figure*}[ht]
 \begin{centering}
   \includegraphics[width=16cm,height=7cm]{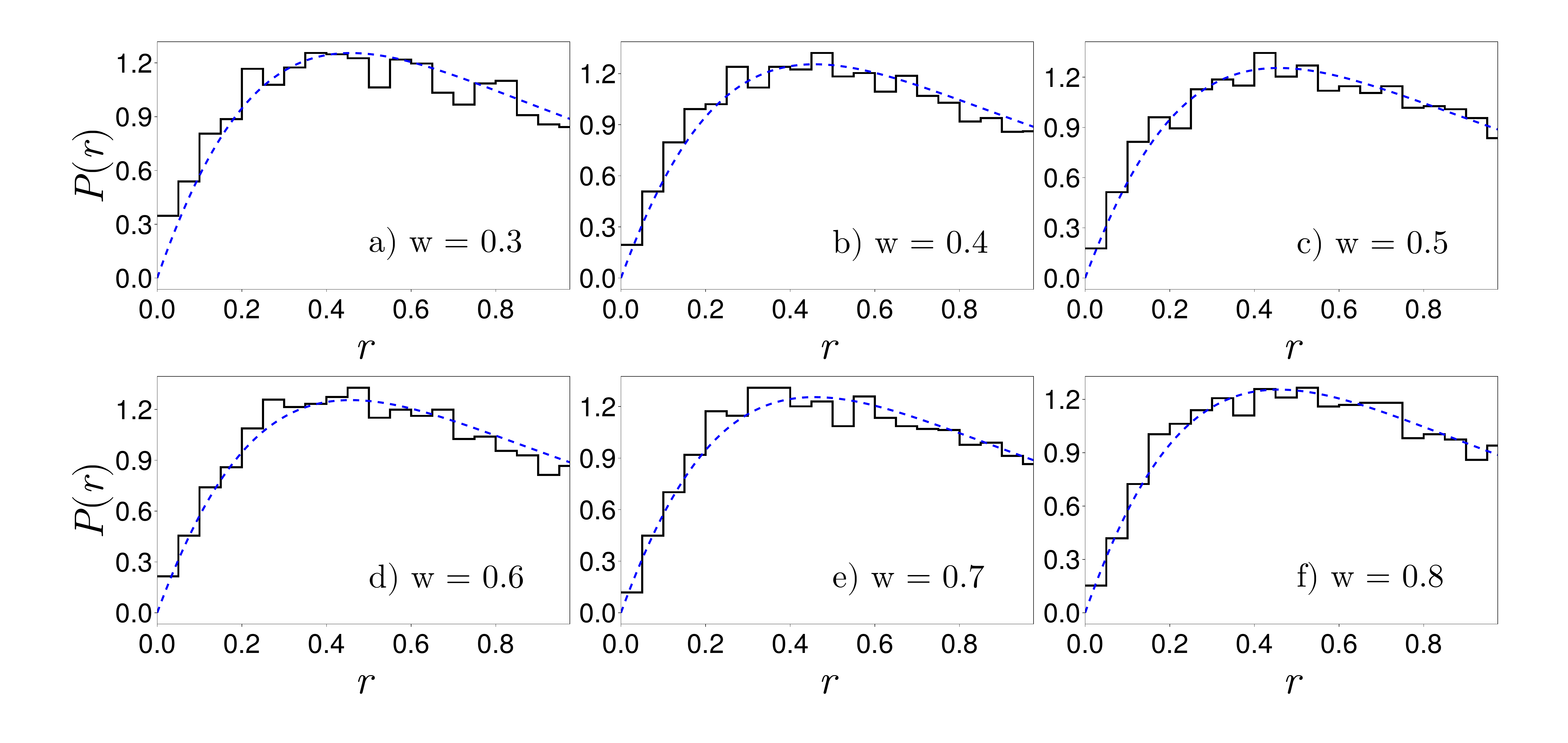}  
   \par\end{centering}
 \caption{Numerical level spacings ratio distribution for the chaotic
   states in the interval $k \in \lbrack 970, 1000 \rbrack $
   compared with the GOE distribution,
   after the separation using the classical criterion,
   for 6 geometries $w=0.3-0.8$ in equal steps of $0.1$,
   from (a) to (f).  The small
   difference demonstrates weak localization effects.}
   \label{figPRchaotichigh}
\end{figure*}

\begin{figure*}[ht]
 \begin{centering}
   \includegraphics[width=16cm,height=7cm]{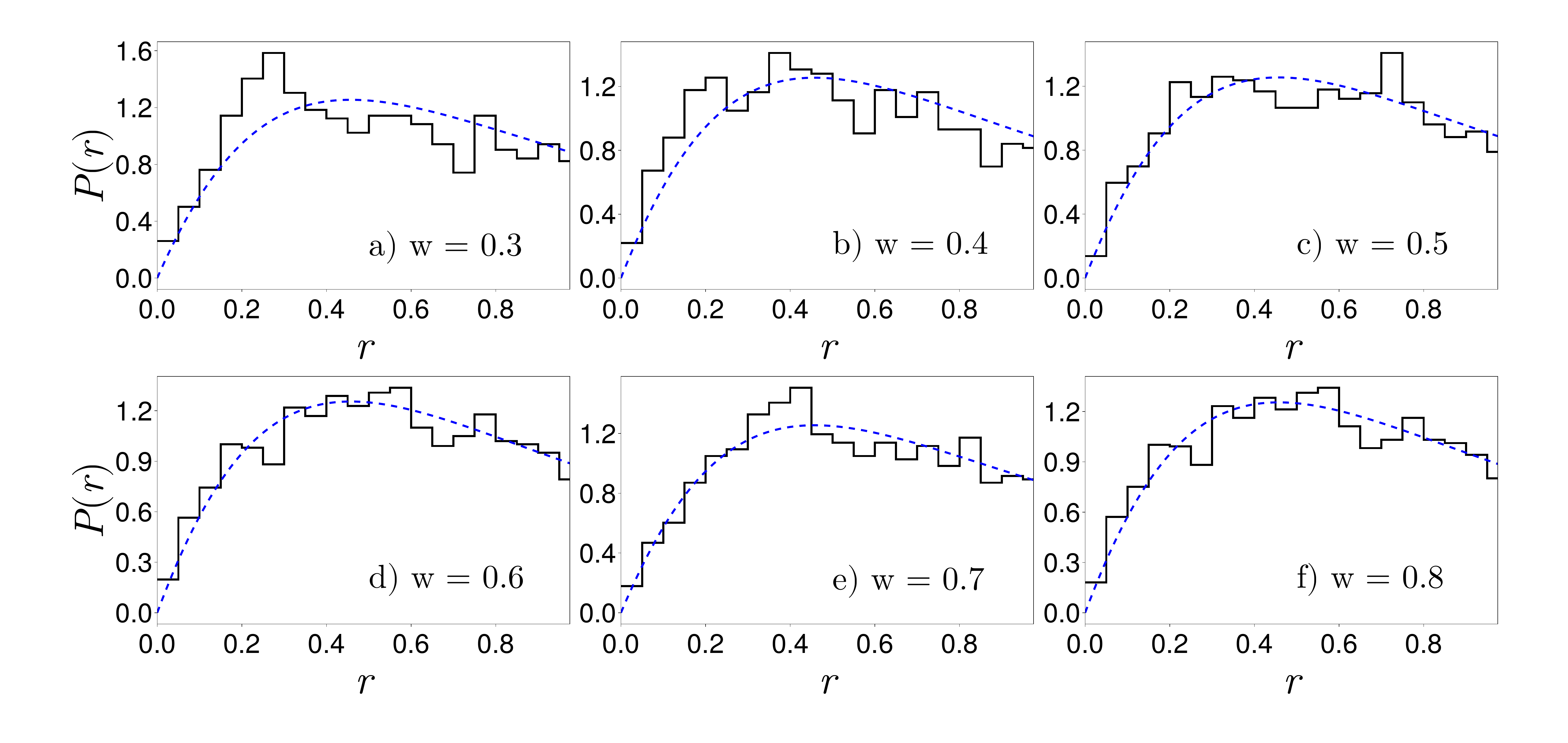}  
   \par\end{centering}
 \caption{The same as in Fig.\ref{figPRchaotichigh}, but for $k\in \lbrack 5, 150\rbrack$,
   showing that even at low energies $k^2$ the effects of localization also are minor,
   but definitely larger than at high-lying states $k \in \lbrack 970, 1000 \rbrack $
   of Fig.\ref{figPRchaotichigh}.}
\label{figPRchaoticlow}
\end{figure*}

\section{Classical Separation Criterion}
\label{appB}

The central idea of this algorithm is to use an overlap grid $C_{q(i),p(j)}$ defined in the classical phase space as having values $+1$ on the chaotic component and $-1$ on the regular component. An example of such a grid is given in Fig.\ref{figHusimi} below. We then project this grid onto the normalized Poincaré - Husimi function $H(q(i),p(j))$ and calculate it's overlap as $M = \sum_{i,j}C_{i,j}H_{i,j}$ where $(i,j) \in N \times N$ are the grid indexes common to both $M$ and $C$. This procedure is repeated for every eigenstate in the chosen interval of study. If we were deep enough into the semiclassical limit we would only be left with states that have the value of $M=+ 1$ for chaotic states and $M = -1$ for regular ones, but before reaching that limit we will usually have a mixture of values $M \in (-1,1)$. So to determine which states are chaotic we choose a threshold value $M_{th}<1$ and calculate the fraction of states that have their $M$ index larger than it. We compare this fraction to the chaotic (3D) phase-space volume for that geometry (Fig.\ref{figmuc}). If the value is higher than the theoretical value we raise the value of $M_{th}$ and vice-versa. We repeat this process until we are close enough to the theoretical classical value (usually $<2 \%$ relative closeness).

\section{The distribution of spacing ratios for the chaotic eigenstates in the low and high regions}
\label{appC}

In Figs. \ref{figPRchaotichigh} and \ref{figPRchaoticlow}, we show the distribution of spacing ratios for chaotic eigenstates, selected according to the classical separation criterion. Compared to the localization evident in the level spacing distributions shown in Figs. \ref{figPSchaotichigh} and \ref{figPSchaoticlow}—where the Brody parameter $\beta$ serves as an indicator of non-ergodic extended states or localization among chaotic eigenstates—the spacing ratio distributions exhibit no clear deviation from the Wigner surmise result given in Eq. \eqref{PrPoissonWigner}, for large Brody parameters. An analytical expression for the spacing ratio distribution corresponding to the Brody distribution in level spacings is essential for extracting localization phenomena from spacing ratios; however, this remains an open problem.




\bibliography{or.bib}

\end{document}